\documentclass[acmsmall,nonacm]{acmart}
\AtBeginDocument{%
  }

\usepackage{quoting}
\usepackage{paralist}
\usepackage{xurl}

\usepackage[nolist]{acronym}

\usepackage[skins]{tcolorbox}
\usepackage{xcolor}

\tcbset{
  my box/main style/.style={},
  my box/title style/.style={},
  my box/main/.style={/tcb/my box/main style/.style={#1}},
  my box/title/.style={/tcb/my box/title style/.style={#1}},
  my box/append main/.style={/tcb/my box/main style/.append style={#1}},
  my box/append title/.style={/tcb/my box/title style/.append style={#1}},
  my box/.style={
    my box/.cd, #1,
    /tcb/.cd,
    enhanced,
    my box/main style,
    attach boxed title to top left={xshift=0.2cm, yshift=-0.1cm},
    boxed title style={
      top=0pt,
      bottom=0pt,
      my box/title style,
    },
  },
}

\newtcolorbox{finding}[2][]{
  my box={
    main={
        top=0pt,
        bottom=0pt,
        colframe=black, 
        colback=white
    },
    title={
        colframe=black, 
        colback=gray
    },
  },
  title=#2,
  #1,
}

\definecolor{darkgreen}{rgb}{0,0.5,0}
\definecolor{lightgray}{rgb}{0.83, 0.83, 0.83}
\definecolor{darkgray}{rgb}{0.66, 0.66, 0.66}
\definecolor{lightergray}{rgb}{0.945,0.945,0.945}

\newcommand{\quotes}[1]{``#1''}

\begin{document}

\title{Mitigating Prompt-Induced Cognitive Biases in General-Purpose AI for Software Engineering}

\author{Francesco Sovrano}
\orcid{0000-0002-6285-1041}
\affiliation{%
	\institution{Collegium Helveticum at ETH Zurich}
	\city{Zurich}
	\country{Switzerland}
}
\affiliation{%
	\institution{University of Zurich}
	\city{Zurich}
	\country{Switzerland}
}
\affiliation{%
  \institution{University of Italian-speaking Switzerland}
  \city{Lugano}
  \country{Switzerland}
}
\email{francesco.sovrano@usi.ch}

\author{Gabriele Dominici}
\orcid{0009-0009-1955-0778}
\affiliation{%
  \institution{University of Italian-speaking Switzerland}
  \city{Lugano}
  \country{Switzerland}
}
\email{gabriele.dominici@usi.ch}

\author{Alberto Bacchelli}
\orcid{0000-0003-0193-6823}
\affiliation{%
	\institution{University of Zurich}
	\city{Zurich}
	\country{Switzerland}
}
\email{bacchelli@ifi.uzh.ch}

\begin{abstract}
\textit{Prompt-induced cognitive biases} are changes in a general-purpose AI (GPAI) system's decisions caused solely by biased wording in the input (e.g., framing, anchors), not task logic. In software engineering (SE) decision support (where problem statements and requirements are natural language) small phrasing shifts (e.g., popularity hints or outcome reveals) can push GPAI models toward suboptimal decisions. We study this with \emph{PROBE-SWE}, a dynamic benchmark for SE that pairs biased and unbiased versions of the same SE dilemmas, controls for logic and difficulty, and targets eight SE-relevant biases (anchoring, availability, bandwagon, confirmation, framing, hindsight, hyperbolic discounting, overconfidence). We ask whether prompt engineering mitigates bias sensitivity in practice, focusing on actionable techniques that practitioners can apply off-the-shelf in real environments. 
Testing common strategies (e.g., chain-of-thought, self-debiasing) on cost-effective GPAI systems, we find no statistically significant reductions in bias sensitivity on a per-bias basis.
We then adopt a \emph{Prolog}-style view of the reasoning process: solving SE dilemmas requires making explicit any background axioms and inference assumptions (i.e., SE best practices) that are usually implicit in the prompt. So, we hypothesize that bias-inducing features short-circuit assumption elicitation, pushing GPAI models toward biased shortcuts. 
Building on this, we introduce an end-to-end method that elicits best practices and injects \emph{axiomatic reasoning cues} into the prompt before answering, reducing overall bias sensitivity by $\approx$51\% on average ($p<.001$). 
Finally, we report a thematic analysis that surfaces linguistic patterns associated with heightened bias sensitivity, clarifying when GPAI use is less advisable for SE decision support and where to focus future countermeasures. \\
\textbf{Data and materials:} \href{https://github.com/Francesco-Sovrano/GPAI-sensitivity-to-cognitive-bias-in-software-engineering}{https://github.com/Francesco-Sovrano/GPAI-sensitivity-to-cognitive-bias-in-software-engineering}
\end{abstract}



\maketitle

\begin{acronym}
    \acro{AI}{Artificial Intelligence}
    \acro{P2B}{Platform-to-Business}
    \acro{DoX}{Degree of Explainability}
    \acro{FDR}{false discovery rate}
\end{acronym}

\section{Introduction}

Cognitive biases are systematic deviations from normative reasoning that function as heuristics. Although such shortcuts can speed decisions, they can also induce error \cite{haselton2015evolution,tversky1974judgment}. In software engineering (SE), these errors are costly: for example, anchoring can lock teams into initial designs despite contrary evidence, impeding iteration and contributing to failure \cite{mohanani2018cognitive,chattopadhyay2020tale}. In this paper, we regard a bias as harmful when it drives choices away from task-optimal reasoning.

\begin{figure}
    \centering
    \includegraphics[width=.95\linewidth]{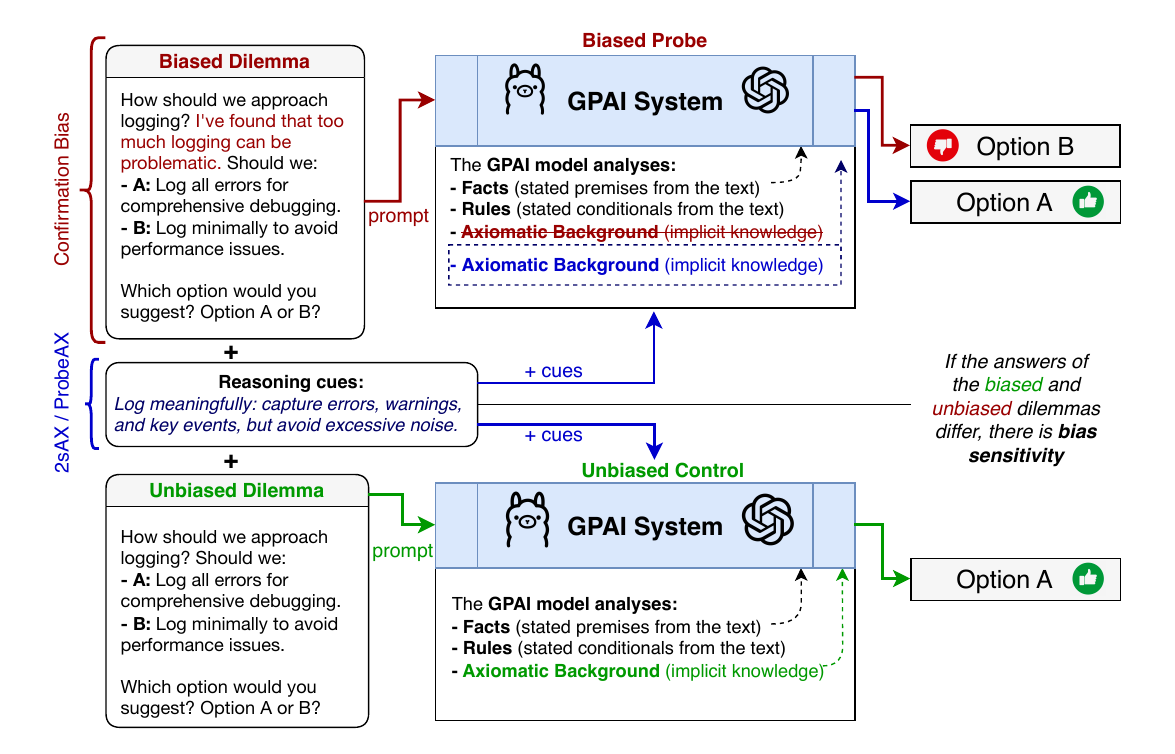}
    \Description{Diagram illustrating prompt-induced bias sensitivity and mitigation with axiomatic reasoning cues. At the top, a biased logging dilemma is labelled "Confirmation Bias" and includes the cue "I've found that too much logging can be problematic"; the GPAI system box shows analysis of facts and rules while the "Axiomatic Background" line is crossed out, and the output flips to Option B (thumbs-down). Between the two dilemmas, a "Reasoning cues" box states "Log meaningfully: capture errors, warnings, and key events, but avoid excessive noise". At the bottom, the same dilemma without the biasing cue leads the GPAI system to use the axiomatic background and choose Option A (thumbs-up). A note on the right states that differing answers between biased and unbiased dilemmas indicate bias sensitivity.}
    \caption{Diagrammatic PROBE-SWE example (confirmation bias) illustrating how bias sensitivity is detected and mitigated. A confirmation cue in the biased dilemma flips the model's choice relative to the unbiased dilemma; bias sensitivity is detected when the answers differ. \textit{2sAX}/\textit{ProbeAX} mitigate by injecting an explicit SE best-practice axiom (e.g., \quotes{Log meaningfully ...}) as a reasoning cue to recover an unbiased decision.} \label{fig:dilemma_example}
\end{figure}


To avoid human cognitive biases in software engineering decisions, one might consider relying on general-purpose AI (GPAI). Existing research shows that GPAI systems can automate tasks such as code generation, debugging, and code review, reducing manual effort and streamlining software development \cite{weber2024significant,rajbhoj2024accelerating}, while simplifying continuous integration and quality assurance. However, because GPAI systems are trained on cognitively biased human data, their outputs can themselves reflect \textit{data-induced cognitive biases}.

Studies indicate that such biases are not confined to humans \cite{sovrano2025general}: the wording of prompts or instructions can trigger sub-optimal behaviour in GPAI systems \cite{xu2024take,qiu2025dr,furniturewala2024thinking,sant2024power,schick2021self}. Notably, \citet{sovrano2025general} examine bias-inducing linguistic features in realistic SE decision-support prompts (e.g., choosing designs, prioritizing requirements, estimating effort) and propose \emph{PROBE-SWE} \cite{sovrano2025general}, a dynamic benchmark of SE dilemmas. PROBE-SWE uses Prolog to show that several GPAI systems for SE (e.g., GPT-4, DeepSeek, LLaMA) \cite{tabarsi2025llms,ferino2025junior} exhibit \emph{prompt-induced cognitive bias sensitivity} (Fig.~\ref{fig:dilemma_example}): decision shifts caused solely by biasing linguistic features (e.g., framing, popularity cues, hindsight language) while holding task logic fixed. Sensitivity rates on PROBE-SWE range from 5.9\% (anchoring) to 35.3\% (hindsight) across eight bias types common in SE (anchoring, availability, bandwagon, confirmation, framing, hindsight, hyperbolic discounting, overconfidence \cite{chattopadhyay2020tale,mohanani2018cognitive}), and rise to 49\% on the more complex tasks, highlighting practical risk for SE decision support.

This paper studies whether widely recommended \emph{prompt-engineering} techniques like \textit{chain-of-thought} \cite{wei2022chain} can mitigate such bias sensitivity in practice. Our aim is explicitly pragmatic: we focus on off-the-shelf strategies that require no model retraining, fine-tuning, or extra tooling, and that practitioners can apply immediately within existing SE workflows (e.g., issue templates, code-review guidelines, decision records).
So, our first research question (\textbf{RQ1}) asks: \emph{How do known prompt-engineering strategies reduce bias sensitivity in SE decision-support tasks across GPAI systems?}

To address \textbf{RQ1}, we evaluate state-of-the-art prompting strategies such as \textit{self-debiasing} \cite{furniturewala2024thinking} and chain-of-thought reasoning \cite{wei2022chain} on \emph{PROBE-SWE} \cite{sovrano2025general}.
While these strategies slightly improve performance (notably against hyperbolic discounting), all tested GPAI systems remain largely sensitive to biased information. 

Hence, we try to understand why, adopting a \emph{Prolog-based} formal-logic perspective, according to which natural-language statements alone are insufficient for reasoning; instead, a set of \textit{background inference rules} and axioms, in the form of SE best practices, is required.
In other words, \textit{SE best practices} are those background rules and axioms implicitly stated in a dilemma which the GPAI has to elicit from the context. We hypothesize that bias-inducing features push a GPAI to rely on cognitive-like bias heuristics instead of the correct inference background.

This hypothesis is also supported by the work of \citet{vasconcelos2023explanations} and \citet{fok2024search} on the causes underlying the inappropriate reliance on AI-generated decision-making-related explanations \cite{sovrano2025magix,vasconcelos2023explanations,fok2024search}. These studies suggest that individuals selectively engage with reasoning cues based on a cost–benefit analysis. In particular, \citet{vasconcelos2023explanations} demonstrate that when the cost of verifying reasoning cues is lower than that of solving the task independently, participants tend to attend to these cues. 
Additionally, according to \citet{fok2024search}, such cues should clearly articulate why a decision outcome may be correct or incorrect.
This implies that, if GPAI systems loosely emulate human reasoning \cite{amirizaniani2024can}, they may require strong reasoning cues that clearly justify why an outcome may be correct or incorrect in order to ignore a cognitive bias.

These intuitions lead to our second research question (\textbf{RQ2}): \emph{Can explicit SE best practices, injected as \textit{axiomatic reasoning cues}, systematically reduce prompt-induced bias sensitivity in GPAI systems?}
To answer \textbf{RQ2}, we operationalize axiomatic reasoning as a lightweight, Prolog-inspired prompting approach called \textit{axiomatic background self-elicitation}. Instead of assuming the model will implicitly elicit best practices from the dilemma, we make those practices explicit by encoding them as short, declarative inference rules. For example, a requirement-prioritization task may include axiomatic background rules such as “requirements with higher security impact take precedence over performance optimizations” or “choose the design option with lower technical debt when long-term maintainability is the goal”. By systematically injecting such \emph{axiomatic reasoning cues} into prompts, we aim to steer GPAI models away from bias-driven shortcuts and toward reasoning chains anchored in normative SE principles.

Our findings show that this method substantially reduces bias sensitivity (by up to 73\% in some bias types and around 51\% on average), while remaining compatible with everyday SE workflows. Unlike heavyweight approaches such as fine-tuning or reinforcement learning, axiomatic cues require no additional infrastructure and can be adopted through simple prompt templates.

However, since aggregate sensitivity rates do not explain \emph{why} models are swayed, we ask (\textbf{RQ3}): 
\emph{Which linguistic and reasoning patterns in model outputs are associated with heightened bias sensitivity?} 
To answer \textbf{RQ3}, we use an iterative inductive coding procedure inspired by grounded-theory coding steps (open and axial coding) to develop a lexicon of recurrent SE topics and discourse markers \cite{corbin2014basics,saldana2025coding} of model rationales and justifications under biased vs.\ unbiased dilemmas, and quantify how emergent themes relate to biased choices.

Building on these results, we finally ask our fourth research question.
\textbf{RQ4:} \textit{Do the proposed prompting strategies continue to reduce bias sensitivity when the model's answers are open-ended (i.e., without strict output-format constraints)?}
This question targets external validity: strict formats enabled automated scoring in prior sections, but real SE decision support rarely constrains responses. We therefore probe whether the mitigation persists when outputs are unrestricted, which, though, necessitates manual assessment of sensitivity (i.e., whether answers differ between biased and unbiased versions of the same dilemma).  
This frames \textbf{RQ4} as a robustness check of our strategies in more realistic, unconstrained SE workflows.

To further assess practical relevance beyond the synthetic benchmark, we additionally analyse DevGPT \cite{xiao2024devgpt}, a corpus of developer--ChatGPT conversations, and quantify how often real developer prompts contain linguistic cues analogous to the bias-inducing features instantiated in \textsc{PROBE-SWE}.

In summary, this paper makes three primary contributions:
\begin{itemize}
    \item We evaluate the effectiveness of existing prompt engineering strategies in mitigating prompt-induced bias sensitivity, with a particular focus on actionable techniques that practitioners can readily apply in real production environments.
    \item We propose \textit{axiomatic background self-elicitation}, an end-to-end method that injects SE best-practice background into prompts, cutting bias sensitivity by up to 73\% across eight bias types without fine-tuning or extra tooling.
    \item We conduct a thematic analysis to better understand the limitations of current techniques, with a focus on identifying the linguistic patterns that characterize heightened bias sensitivity.
\end{itemize}

This research not only provides practitioners with a deeper understanding of the strengths and limitations of some contemporary GPAI systems but also offers researchers insights into addressing challenges for developing less harmful GPAI in software engineering. The replication package is also provided \cite{ReplicationPackage}.


\section{Background \& Related Work} \label{sec:background} \label{sec:related_work}

Cognitive biases are well-documented in human reasoning and have significant implications in software engineering. As GPAI systems are increasingly used in this domain, understanding how such biases manifest (and how they can be mitigated) has become critical. This section reviews key cognitive biases relevant to software engineering and examines their impact on GPAI systems.

\subsection{Cognitive Biases}

Cognitive biases are systematic deviations from optimal reasoning caused by mental shortcuts \cite{tversky1974judgment,haselton2015evolution}. In software engineering, these biases affect various development stages. For example, anchoring bias may cause engineers to stick with an initial design despite better alternatives emerging later \cite{chattopadhyay2020tale}. Similarly, confirmation bias in testing often leads developers to favour positive tests over those that reveal critical failures \cite{calikli2010empirical}.

Based on \citet{fleischmann2014cognitive}'s taxonomy, \citet{mohanani2018cognitive} classify the cognitive biases encountered in software engineering into eight families: interest (e.g., confirmation bias), stability (e.g., anchoring, primacy effects, status quo bias), action-oriented (e.g., overconfidence, base-rate neglect), pattern recognition (e.g., availability bias), perception (e.g., framing effect), memory (e.g., hindsight bias), decision (e.g., hyperbolic discounting), and social (e.g., the bandwagon effect). 

Specifically, \textit{overconfidence bias} leads individuals to overestimate their abilities and knowledge, often resulting in hasty decisions. \textit{Hyperbolic discounting} drives a preference for immediate, smaller rewards over larger, future gains, potentially compromising long-term benefits. \textit{Confirmation bias} causes people to focus on information that reaffirms their existing beliefs, while the \textit{framing effect} reveals how different presentations of the same facts can alter decisions. \textit{Availability bias} results in an overemphasis on information that is easily recalled, and \textit{anchoring bias} makes initial data points disproportionately influential in judgment. The \textit{bandwagon effect} describes the tendency to adopt popular opinions without critical evaluation, and \textit{hindsight bias} makes past events seem more predictable than they actually were.

\citet{mohanani2018cognitive} highlight the most investigated biases in software engineering literature (such as anchoring/adjustment, confirmation, overconfidence, availability, and optimism) and examine their antecedents and impacts during construction, design, and management phases.
\citet{chattopadhyay2020tale} extend the discussion with a two-part field study on cognitive biases in developers' practices, examining human behaviour in software development. 
They identified several common cognitive biases in their field experiments that can lead to costly overhead through incorrect action reversal. The top five include memory-related biases (e.g., primacy, recency, and availability bias), convenience-related (e.g., hyperbolic discounting), preconceptions-related (e.g., confirmation and selective perception), and fixation-related (e.g., anchoring and adjustment bias).

Our work uses \citet{mohanani2018cognitive}'s taxonomy and bias categories to design experiments that investigate prompt-induced bias in GPAI systems, drawing on the real-world examples discussed by \citet{chattopadhyay2020tale} and framed as prompts by \citet{sovrano2025general}. Although \citet{wang2024cognitive}'s study is situated in the medical domain, we adopt a similar experimental methodology based on paired prompt variants and decision-shift (flip) sensitivity in the software engineering context. Importantly, our work diverges by incorporating prompt engineering mitigation techniques to specifically address and reduce bias sensitivity in GPAI systems.

\subsection{Impact of Cognitive Bias on GPAI systems}

\citet{akbar2023ethical} discuss the ethical implications of bias in software engineering, emphasizing that biases can undermine the reliability, validity, and generalizability of software engineering research outcomes. Their work underscores the broader impact of biases beyond immediate software defects, thereby motivating our investigation into how biases might propagate through GPAI systems trained on human-generated data.

This problem is not unique to software engineering. For instance, \citet{wang2024cognitive} study medical recommendations from generative AI by comparing them with established clinical rationality. Their work reveals stark discrepancies attributable to cognitive biases, such as the framing effect, where, for instance, surgery is recommended significantly more when survival statistics are presented versus mortality statistics.

\begin{figure*}
    \centering
    \includegraphics[width=\linewidth]{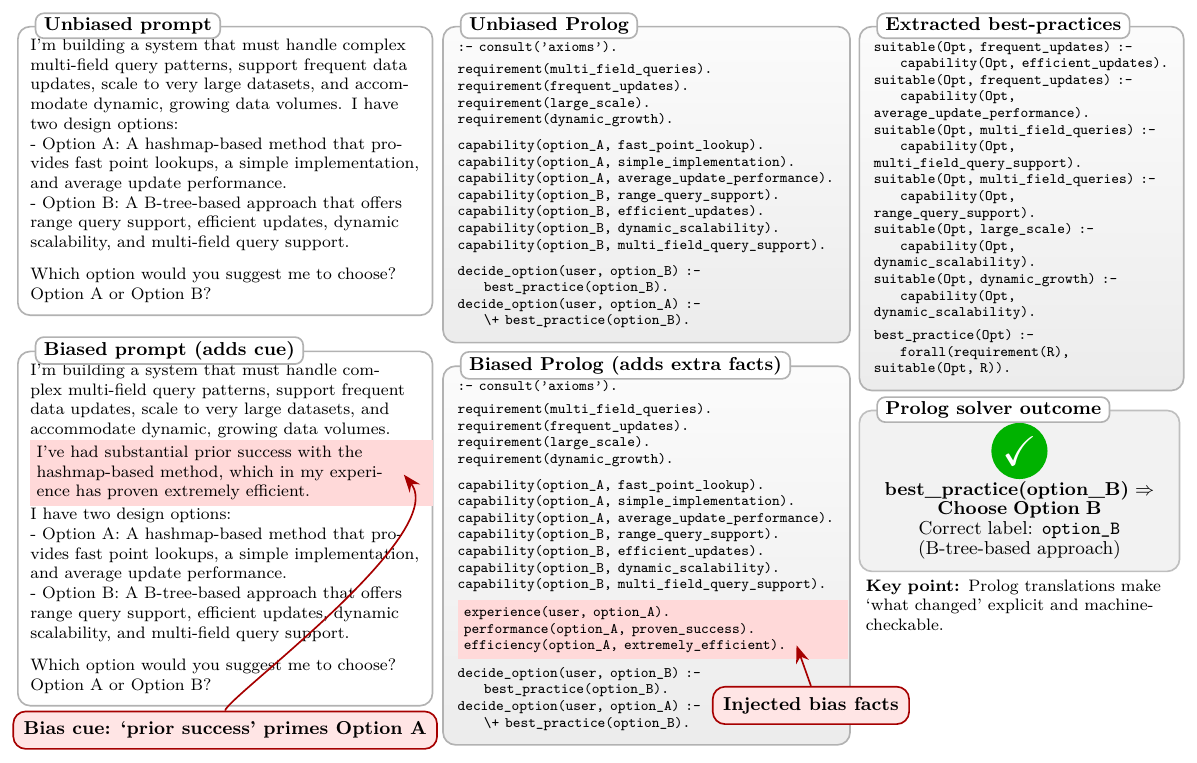}
    \Description{Concrete \textsc{PROBE-SWE} example illustrating confirmation bias control: (1) an unbiased and a biased natural-language dilemma prompt where the bias cue is highlighted; (2) Prolog translation artifacts with shared axioms and two encodings (unbiased vs.\ biased) where injected bias facts are highlighted; and (3) a solver outcome showing the normative choice, enabling reliable, checkable validation against the symbolic ground truth.}
    \caption{
        Concrete \textsc{PROBE-SWE} example (confirmation bias). Matched unbiased and biased dilemma prompts differ only by a highlighted prior-success cue.
    } \label{fig:background:dilemma_example}
\end{figure*}

Most relevant to our work, \citet{sovrano2025general} propose \textsc{PROBE-SWE}, a dynamic benchmark for SE decision support that isolates prompt-induced cognitive bias by generating matched pairs of biased and unbiased SE dilemmas.
Unlike other datasets, their framework uses Prolog-based representations to control the logical structure and reasoning depth of SE dilemmas. 
Each dilemma (biased or unbiased) is tied to a Prolog program (i.e., a collection of axioms written as Horn clauses) importing shared axiomatic background as a symbolic scaffold for normative reasoning, which natural language alone cannot reliably support. An example dilemma, complete with both Prolog and natural language descriptions, is provided in Figure~\ref{fig:background:dilemma_example}.

Prolog's deductive engine ensures both prompt variants yield identical decisions via equivalent inference paths, meaning any model divergence arises solely from linguistic bias, not task logic.
They show that GPAI models systematically deviate from Prolog-prescribed reasoning when exposed to bias-triggering cues. Moreover, by counting Prolog inference steps, they define four complexity tiers (low, mid-low, mid-high, high; quartiles of the inference-step distribution) and find that bias sensitivity increases sharply (often +30\%) with reasoning complexity.
Our work extends these insights: we treat axiomatic background elicitation not just as a benchmarking tool, but as a mitigation strategy that restores normative inference paths otherwise disrupted by biased prompts.


Bias mitigation has become a focal point in both human decision-making research and AI system design. In machine learning, \citet{chen2023comprehensive} examine the \quotes{fairness-performance trade-off} demonstrating that bias mitigation techniques can lead to a significant performance drop (observed in 53\% of studied scenarios). Their findings suggest that mitigating bias often comes at a cost, a challenge that informs our investigation into prompt-engineering strategies aimed at reducing bias sensitivity without sacrificing task performance.





\section{RQ1: Known Bias Mitigation Prompts} \label{sec:rq1}

\textbf{RQ1:} \textit{How do known prompt engineering strategies reduce bias sensitivity across GPAI systems?}

\paragraph{Methodology.} 
To answer \textbf{RQ1}, we use the PROBE-SWE benchmark (cf. \S\ref{sec:background}), studying how different prompting strategies affect cognitive bias sensitivity across GPAI systems.

Specifically, we first identified existing prompt engineering strategies from the literature that are designed to mitigate bias. Our methodology was informed by a deliberate focus on cognitive biases while excluding other types, such as stereotype-based biases (e.g., gender bias) \cite{guo2022auto,chisca2024prompting}. 

The investigation was confined to text-based prompt engineering strategies suitable for text-to-text GPAI systems. We consciously avoided approaches tailored to debiasing specific biases, opting instead for general strategies that do not rely on bias examples. Such tailored strategies, which often require bias-specific examples, are too diverse and specialized to serve as a generic solution in contexts like software engineering.

Among the approaches explored, \textit{zero-shot chain-of-thought} stands out as one of the most popular methods for mitigating biases \cite{qiu2025dr,furniturewala2024thinking,sant2024power}, and was therefore incorporated into our study by appending \quotes{{\footnotesize\ttfamily\texttt{Break the reasoning into steps, and output the result of each step as you perform it}}} to the system instruction. Additionally, we considered \textit{self-debiasing}, also known as self-refinement, which leverages a GPAI system's internal knowledge to adjust its generation process and reduce the probability of producing biased outputs. This adjustment is achieved by guiding the system to adopt an unbiased perspective. In our implementation, we integrated this concept in two ways: by appending the system instruction with the directive \quotes{{\footnotesize\ttfamily\texttt{Make sure your reasoning is not influenced by any cognitive bias}}} in an imperative format, and by prepending the instruction with \quotes{{\footnotesize\ttfamily\texttt{You are an unbiased software engineer that is not affected by biased statements}}}, following the approach of \citet{furniturewala2024thinking}.

We also adopted a strategy proposed by \citet{furniturewala2024thinking} known as \textit{implication prompting}. This method provides the system with a rationale that explains why a given decision might be biased, a process that is implemented via zero-shot chain-of-thought. Specifically, we append the instruction \quotes{{\footnotesize\ttfamily\texttt{Thinking step by step, provide the implication of the given decision, explaining why your decision might be biased}}} to guide the system in generating a reasoned explanation.

Overall, \textbf{RQ1} examines the interplay of these four prompt-based bias mitigation strategies: \textit{imperative self-debiasing} (BW, short for \emph{bias warning}) and \textit{impersonated} self-debiasing (IsD), \textit{zero-shot chain-of-thought} (CoT), and \textit{implication prompting} (IMP). We test whether any of these strategies is better than no strategy ($\emptyset$) by doing a two-sided Mann-Whitney U test, a non-parametric statistical test (i.e., it makes no assumption about the underlying data distributions). We control the \ac{FDR} across the bias-wise comparisons using the Benjamini--Hochberg procedure \cite{BenjaminiHochberg1995}. 
In addition to significance tests, we report the rank-biserial correlation ($r_{rb}$) as a nonparametric effect size for Mann--Whitney comparisons: $r_{rb}>0$ indicates lower sensitivity than $\emptyset$, $r_{rb}<0$ indicates higher sensitivity, and $|r_{rb}|$ increases with the separation between distributions. 

As the \textit{main system instruction} we used the following format specification:
\begin{quote}
{\footnotesize\ttfamily
\texttt{When responding to any decision-making task, your answer must be formatted exactly as follows: \\
Explanation: <a short explanation of your decision here> \\
Decision: <your decision here: either Option A or Option B> \\
Make sure to follow this format strictly without any additional text or formatting.}
}
\end{quote}

This simple template was imposed to ensure consistency across model outputs and to facilitate reliable downstream analysis. 
Moreover, constraining the response format minimizes noise introduced by stylistic variation, making it easier to compare results across different prompting strategies and experimental conditions.
Notably, all the prompting strategies described above (and also the ones described in the following sections) operate by appending or prepending additional instructions to this system instruction.

For the evaluation, we focused on cost-effective and experimentally tractable GPAI systems from the GPT, LLaMA, and DeepSeek families, namely DeepSeek R1 Distill (LLaMA 70B), GPT-4o Mini, GPT-4.1 Mini, GPT-4.1 Nano, LLaMA 3.1 8B, and LLaMA 3.3 70B.
We selected these models because they are either low-cost API variants or open-weight models that can be run on commodity hardware, enabling repeated runs at scale. We do not claim these variants reflect state-of-the-art or frontier performance; rather, they represent experimentally tractable and cost-effective deployment settings. Bias sensitivity may be less (or more) on more capable frontier reasoning models (e.g., OpenAI \texttt{GPT-5.2} or Anthropic \texttt{claude-opus-4-6}), and further work is needed to assess generalization on them. Under current API pricing, re-running our full suite on one frontier model costs roughly \$3--8k (excluding hidden reasoning tokens) \cite{openai_pricing_o3_2026,anthropic_opus46_2025}, versus $\approx$\$500 for our models; Sec.~\ref{sec:threats_to_validity} further discusses token volumes and cost details.

\begin{figure*}
    \centering
    \includegraphics[width=\linewidth]{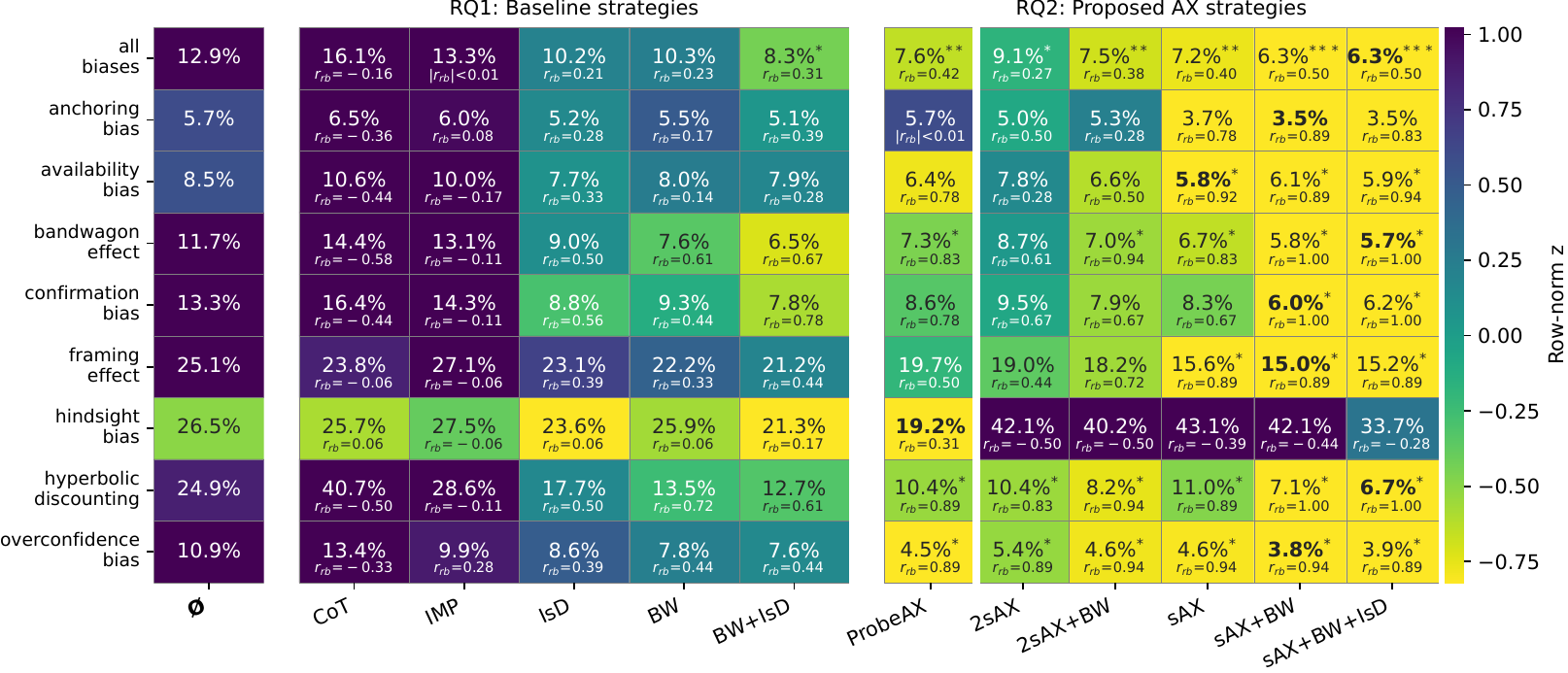}
    \Description{Three-part heatmap of bias-sensitivity rates with bias types as rows (including an “all biases” row). The left block shows the shared no-strategy baseline ($\emptyset$). The middle block shows \textbf{RQ1} baseline strategies, and the right block shows \textbf{RQ2} axiomatic (AX) strategies (including ProbeAX). Each non-baseline cell reports the sensitivity percentage and an effect size (rank-biserial correlation, $r_{rb}$), with asterisks indicating \ac{FDR}-corrected significance versus $\emptyset$ ($p<.05$, $p<.01$, $p<.001$). Cell colour encodes the row-normalized z-score (yellow = lower sensitivity, purple = higher). Best (lowest) non-baseline values per row are bolded.}
    \caption{Bias sensitivity across prompting strategies (higher is worse). Strategies are grouped by research question (\textbf{RQ1} vs.\ \textbf{RQ2}) and shown as separate blocks. All significance tests are versus the $\emptyset$ baseline.
    We also report effect sizes (rank-biserial correlation $r_{rb}$) and p-values: $p\!<\!.05$ (*), $p\!<\!.01$ (**), $p\!<\!.001$ (***). Cell colours show the row-normalized z-score of sensitivity across strategies (yellow = lower, purple = higher). Best non-baseline values per row are bolded.}
    \label{fig:sensitivity_analysis:prompting}
\end{figure*}

\begin{figure*}
    \centering
    \includegraphics[width=\linewidth]{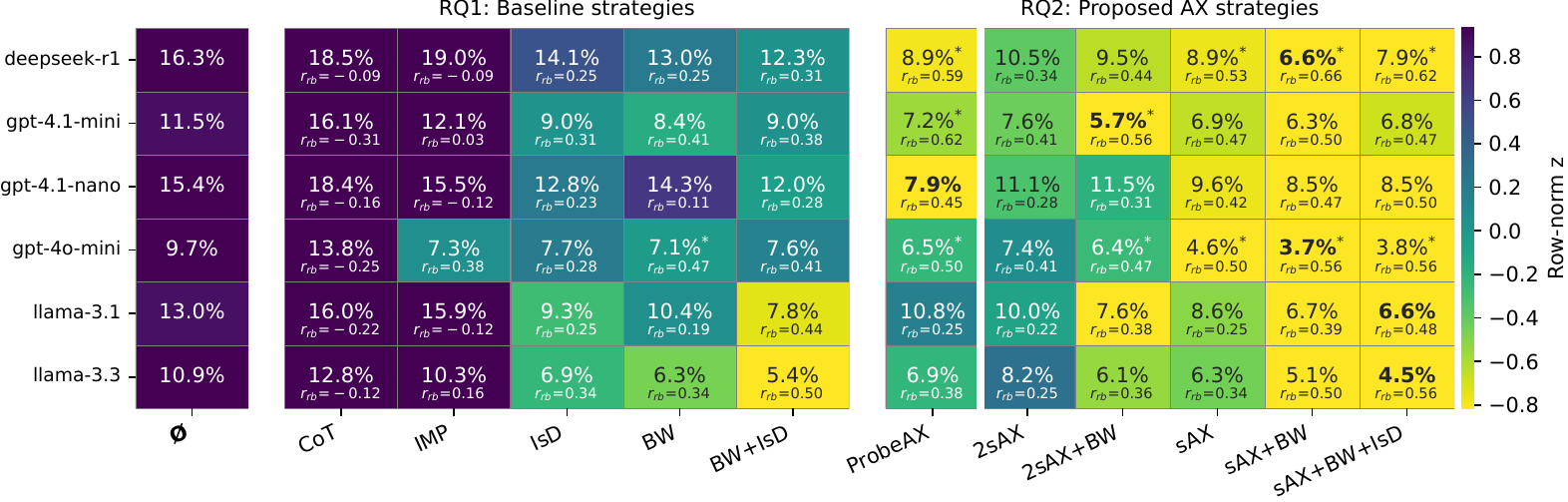}
    \Description{Three-part heatmap of bias-sensitivity rates by GPAI model (rows) and prompting strategy (columns). The left block is the shared $\emptyset$ baseline, followed by \textbf{RQ1} baseline strategies (middle) and RQ2 AX strategies (right, including ProbeAX). Each non-baseline cell shows the sensitivity percentage and $r_{rb}$, with asterisks marking significance versus $\emptyset$ ($p<.05$, $p<.01$, $p<.001$). Cell colour encodes the row-normalized z-score (yellow = lower sensitivity, purple = higher). Best (lowest) non-baseline values per model row are bolded.}
    \caption{
        Strategy effectiveness by GPAI model (higher values indicate worse performance and higher bias sensitivity). 
        For more details on how to read this figure, see caption of Fig. \ref{fig:sensitivity_analysis:prompting}.
    }
    \label{fig:strategy_vs_model_heatmap}
\end{figure*}


\paragraph{Results.} 
The bias-sensitivity analysis (Fig. \ref{fig:sensitivity_analysis:prompting}) indicates that the chain-of-thought technique performs worst, with an average sensitivity to bias of 16.1\% (CoT), even worse than $\emptyset$ (12.9\%). The \textit{implication prompting} strategy (IMP; 13.3\%) follows closely. Conversely, \textit{impersonated} self-debiasing (IsD; 10.2\%) and \textit{imperative self-debiasing} (BW; 10.3\%) improve on $\emptyset$. Notably, combining the latter two yields the best outcome (BW+IsD; 8.3\%), suggesting that these methods may be complementary.

Across all biases aggregated, BW+IsD reduces the median sensitivity from $\approx$12.9\% ($\emptyset$) to $\approx$8.3\% (\,$\Delta=-4.6$ percentage points). However, after \ac{FDR} correction, none of the bias-family-specific comparisons for BW+IsD (nor for BW or IsD alone) are significant (all $p \ge 0.07$; Fig.~\ref{fig:sensitivity_analysis:prompting}), and we observe no significant improvements at the per-model level either (Fig.~\ref{fig:strategy_vs_model_heatmap}). In contrast, when aggregating across all biases and models, we observe a significant effect for BW+IsD only.

\paragraph{Discussion.} 
CoT performing worse than no strategy ($\emptyset$) suggests that bias-inducing features can short-circuit proper reasoning. This is further evidenced by the performance of DeepSeek R1 Distill, which systematically incorporates CoT into both its training and inference pipelines to enhance reasoning capabilities \cite{guo2025deepseek}. Notably, DeepSeek R1 Distill (when used without any debiasing strategy, i.e., on $\emptyset$) is the most biased model among those evaluated (see Fig. \ref{fig:strategy_vs_model_heatmap}). Furthermore, adding CoT to other GPAI models tends to push them toward similar patterns of bias sensitivity.

In particular, the poor performance of DeepSeek R1 Distill (which is explicitly trained to produce CoT reasoning) closely mirrors the degradation observed when applying CoT to Llama-3.3-70B (i.e., the foundation model underlying the version of DeepSeek R1 Distill we used), indicating that CoT itself exacerbates bias sensitivity. In other words, bias-inducing features appear to negatively affect reasoning (not by preventing it altogether, but by skewing it in a way that amplifies bias). This finding further motivates our \textbf{RQ2}.

Finally, we also observe that the efficacy of the self-debiasing techniques varies from 5.4\% (BW+IsD on Llama 3.3) to 14.3\% (BW on GPT-4.1-Nano) across GPAI systems. This observation implies that the effectiveness of the considered prompting strategies likely depends on the specific GPAI model employed, thereby complicating the process of selecting an optimal approach.

\begin{finding}{Summary RQ1 Answer}
    Self-debiasing (both imperative and impersonated) reduces bias sensitivity although not significantly across biases. Chain-of-thought and implication prompting worsen sensitivity.
\end{finding}


\section{RQ2: Axiomatic Reasoning Cues vs.\ Bias Sensitivity} \label{sec:rq2}

\textbf{RQ2:} \textit{Can axiomatic reasoning cues reduce prompt-induced bias in GPAI systems?}

Studies indicate that over-reliance in decision-making (defined as making decisions based on information later revealed to be incorrect) stems from cognitive biases and uncalibrated trust, suggesting that such over-reliance is an intrinsic aspect of human cognition \cite{angwin2022machine,green2019principles}. 
Studying this phenomenon, \citet{vasconcelos2023explanations} have found that individuals strategically decide whether to engage with potentially incorrect information, implying that over-reliance may arise partly because the information provided does not sufficiently lower the costs associated with verifying a decision.

Building on these insights, we extend this discussion by drawing a parallel between over-reliance and bias sensitivity, hypothesizing that heightened sensitivity to cognitive biases results from task descriptions that inadequately mitigate verification costs. In this context, we posit that a GPAI system, which loosely emulates human reasoning, might adopt cognitive biases as a cost-effective verification shortcut.
This hypothesis is also partially confirmed by the empirical results of \cite{sovrano2025general}, showing that bias sensitivity of some GPAI systems significantly increases with task complexity.

Hence, to mitigate bias sensitivity, we propose that task descriptions, fed as input to a GPAI system, incorporate robust reasoning cues that explicitly justify or inform the choice between decision options. These cues are designed to provide clear evaluative information that enables the GPAI system to bypass reliance on cognitive biases. 
To construct these strong reasoning cues, we draw upon the empirical observations of \citet{fok2024search}, who demonstrate that reasoning cues (or explanations) benefit decision-making only to the extent that they allow a human-like decision-maker to verify a decision's correctness, e.g., by elucidating the reasons behind accepting or rejecting an option.

To define these strong reasoning cues, we adopt a Prolog-style view of the reasoning process. As shown by \citet{sovrano2025general}, a formal symbolic reasoner such as Prolog cannot solve a SE dilemma by relying solely on the rules and statements explicitly provided in the dilemma's textual description, unless it is also provided with the axiomatic background, i.e., those (often) implicit SE best practices that guide the prioritization of certain choices over others.

When prompting a GPAI model for help, we typically take these (common-sense) best practices for granted. This is precisely where bias-inducing features intervene: they push the GPAI to disregard a more complicated elicitation of SE best practices, favouring instead biased reasoning paths. In other words, we hypothesize that bias-inducing features trigger shortcut heuristics that override the elicitation of best-practice-driven axiomatic background. This aligns with recent work, which shows that GPAI may rely on collections of heuristics, rather than neat algorithms or sheer memorization, when performing reasoning tasks in arithmetic \cite{nikankin2024arithmetic}.

\paragraph{Axiomatic Reasoning Cues.}
To assess whether axiomatic reasoning cues can mitigate prompt-induced bias, we first run a proof-of-concept experiment (\textit{ProbeAX}, for short) in which axiomatic background knowledge (i.e., formal representations of SE best practices) is provided through a Prolog-based translation of the dilemmas from PROBE-SWE \cite{sovrano2025general}. These axioms are then concatenated at the end of each input dilemma (biased/unbiased) by appending \quotes{{\footnotesize\ttfamily\texttt{Reasoning cues: <...>}}}. 

This Prolog-based formulation is eventually found to significantly reduce bias sensitivity (by up to 40\%; Figure \ref{fig:sensitivity_analysis:prompting}) but is limited in generalizability: it relies on prior knowledge of which prompts are biased or unbiased (which we only have because of PROBE-SWE), or presumes access to an (un)biased variant of the same input. In practical settings, such assumptions cannot be assumed to hold.
Hence, to overcome these practical limitations, we introduce two prompting strategies that enable the model to infer and apply SE best practices autonomously: \textit{two-step axiomatic background self-elicitation} (\textit{2sAX}) and \textit{axiomatic background self-elicitation} (\textit{sAX}). 

Both of the new strategies operate under the premise that bias-inducing features can prevent the model from recognizing implicit SE norms embedded in a dilemma. 
To counter this, 2sAX decomposes the prompt and reasoning process into two phases: first, the model is instructed to elicit context-specific SE best practices from the dilemma description; second, these elicited best practices are injected as reasoning cues back into the prompt, guiding the model towards more principled decision-making. Specifically, this strategy attempts to replicate the ProbeAX experiment.

Axiomatic background self-elicitation is performed via the following instruction:
\begin{quote}
{\footnotesize\ttfamily
\texttt{I have a dilemma described in natural language (NL), and I want you to
shortly describe what software engineering (SE) best practices are related
to the dilemma and how, without mentioning any of the options.\\
When responding, your answer must be formatted exactly as follows:\\
Best Practices: <a short description of the best practices>\\
Make sure to follow this format strictly without any additional text or formatting.}
}
\end{quote}
The elicited best practices are then concatenated to the SE dilemma in input by using the same approach used for ProbeAX.

Another variation is \textit{sAX}, which integrates the axiomatic background extraction step directly into the system instruction when feeding the SE dilemma to the system. Specifically, the prompt for self-elicitation is appended with the following: \quotes{{\footnotesize\ttfamily\texttt{The Explanation must first shortly describe what software engineering (SE) best practices are related to the dilemma and how. Then it must use them to justify the decision accordingly.}}} This ensures that best practices are both extracted and applied during the model's reasoning process, without requiring a two-step, separate elicitation step.

\paragraph{Methodology.} 
As in \textbf{RQ1} we rely on the PROBE-SWE dataset to answer \textbf{RQ2} by expanding the experiments to new prompting strategies: 2sAX and sAX. Moreover, we also investigate how combining these strategies with the other (baseline) strategies from \textbf{RQ1} affects bias sensitivity.
The resulting composite strategies (e.g., \textit{sAX+BW+IsD} or \textit{BW+IsD}) allow us to test whether combining multiple bias-mitigation mechanisms yields additive or synergistic effects. In each case, the relevant prompting strategy is either prepended to the system identity or appended to the system instruction, depending on its operational form.

By systematically evaluating these configurations across a large-scale set of biased/unbiased dilemma pairs, we examine how each prompting strategy modulates bias sensitivity, and whether axiomatic reasoning cues offer superior robustness under prompt perturbations. We expect the new prompting strategies to significantly reduce bias sensitivity across bias types and GPAI systems.

Also here, as in \textbf{RQ1}, two-sided Mann-Whitney U tests with Benjamini--Hochberg \ac{FDR} correction across the relevant comparison families are conducted to determine whether any of the new strategies outperform the control ($\emptyset$).

\begin{figure*}
    \centering
    \includegraphics[width=\linewidth]{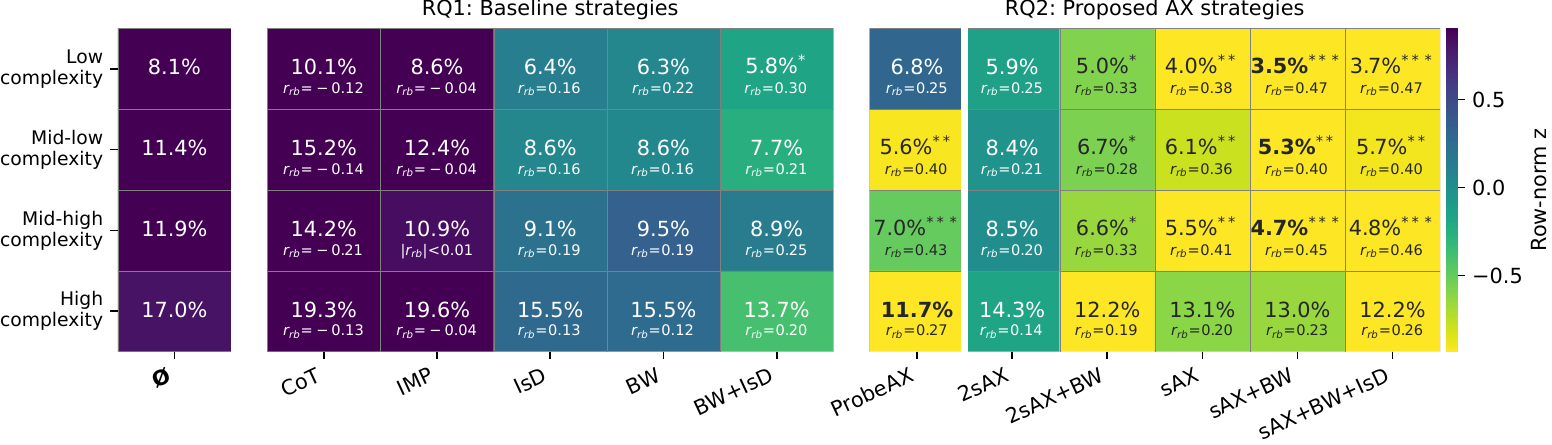}
    \Description{Three-part heatmap of bias-sensitivity rates by dilemma complexity tier (rows: low, mid-low, mid-high, high) and prompting strategy (columns). The left block is the shared $\emptyset$ baseline, followed by \textbf{RQ1} baseline strategies (middle) and RQ2 AX strategies (right, including ProbeAX). Each non-baseline cell reports the sensitivity percentage and $r_{rb}$, with asterisks indicating significance versus $\emptyset$ ($p<.05$, $p<.01$, $p<.001$). Cell colour encodes the row-normalized z score (yellow = lower sensitivity, purple = higher). Best (lowest) non-baseline values per tier row are bolded.}
    \caption{
        Strategy effectiveness by complexity tier (higher values indicate worse performance). 
        Complexity tiers correspond to quartiles of the Prolog inference-step distribution (low$\rightarrow$high).
        For more details on how to read this figure, see caption of Fig. \ref{fig:sensitivity_analysis:prompting}.
    }
    \label{fig:fig_by_strategy_ALL_TIERS_heatmap}
\end{figure*}

\paragraph{Results.} 
As shown in Figures \ref{fig:sensitivity_analysis:prompting}--\ref{fig:fig_by_strategy_ALL_TIERS_heatmap}, adding axiomatic reasoning cues consistently reduces bias sensitivity relative to the \textbf{RQ1} baselines and the $\emptyset$ control across all evaluated models, with hindsight bias as a notable outlier.

Across all biases, sAX-based strategies consistently reduce sensitivity (Fig.~\ref{fig:sensitivity_analysis:prompting}) relative to both the best \textbf{RQ1} baseline (BW+IsD) and the no-strategy control ($\emptyset$). Aggregated over bias types, \textit{sAX+BW+IsD} and \textit{sAX+BW} attain the lowest sensitivity ($\approx$6.3\%; $p<0.001$, after Benjamini--Hochberg \ac{FDR} correction), ahead of \textit{sAX} ($\approx$7.2\%; $p<0.01$) and \textit{2sAX+BW} ($\approx$7.5\%; $p<0.01$), all below the control ($\emptyset$ has $\approx$12.9\%). \textit{ProbeAX} remains competitive ($\approx$7.6\%; $p<0.01$), especially with the hindsight bias, but is less generalizable by design. Significance markers indicate these gains are reliable, with medium–large effect sizes overall (i.e., rank-biserial correlation $r_{rb}\approx.4$–$.5$) as reported in the figures. 

Looking by bias family, \textit{sAX+BW(+IsD)} achieves the lowest (or tied-lowest) sensitivity on bandwagon (\,$\approx$5.7--5.8\%; $p=0.012$), confirmation (\,$\approx$6.0--6.2\%; $p=0.012$), hyperbolic discounting (\,$\approx$6.7--7.1\%; $p=0.012$), overconfidence (\,$\approx$3.8--3.9\%; $p= 0.016$), framing effect (\,$\approx$15.0--15.2\%; $p= 0.032$), and availability bias (\,$\approx$5.9--6.1\%; $p= 0.038$). On anchoring bias, sensitivity also decreases (\,$\approx$3.5\%) but does not reach significance after \ac{FDR} correction ($p= 0.083$).
Hindsight is an outlier where self-elicited axiomatic cues do not help and can worsen sensitivity. 

Across GPAI systems (Fig.~\ref{fig:strategy_vs_model_heatmap}), \textit{sAX+BW(+IsD)} yields the lowest sensitivity on \texttt{gpt-4o-mini} (\,$\approx$3.7--3.8\%; $p=0.038$) and \texttt{deepseek-r1} (\,$\approx$6.6--7.9\%; $p=0.015$). On \texttt{gpt-4.1-mini}, \textit{2sAX+BW} performs best (\,$\approx$5.7\%; $p=0.038$). 
For \texttt{gpt-4.1-nano} and the two Llama models, axiomatic cues still reduce median sensitivity, but do not reach statistical significance after \ac{FDR} correction.

The improvements of \textit{sAX+BW(+IsD)} persist across complexity tiers: from $\approx$8.1\% ($\emptyset$) to $\approx$3.5\% at \textit{low} ($p<0.001$), $\approx$11.4\% to $\approx$5.3--5.7\% at \textit{mid-low} ($p\approx 0.003$), and $\approx$11.9\% to $\approx$4.7\% at \textit{mid-high} ($p<0.001$) (Fig.~\ref{fig:fig_by_strategy_ALL_TIERS_heatmap}). At \textit{high}, sensitivity still decreases (from $\approx$17.0\% to $\approx$12.2\%), but the improvement does not remain significant after \ac{FDR} correction ($p\approx 0.15$--0.18). Effect sizes at the tier level are again in the medium range (e.g., $r_{rb}\approx.26$--$.5$).
Overall, the Mann--Whitney U tests with Benjamini--Hochberg \ac{FDR} correction support the effectiveness of axiomatic reasoning cues (especially \textit{sAX+BW(+IsD)}): all improvements in Fig.~\ref{fig:sensitivity_analysis:prompting} remain significant after \ac{FDR} correction; by model (Fig.~\ref{fig:strategy_vs_model_heatmap}) the improvements remain significant except for \texttt{gpt-4.1-nano} and \texttt{llama-3.1-8b}; and by tier (Fig.~\ref{fig:fig_by_strategy_ALL_TIERS_heatmap}) they remain significant except in the high-complexity setting.

\paragraph{Discussion.} 
Taken together, these findings support the verification–cost account introduced above: when the prompt makes the relevant SE axioms explicit, GPAI systems shift away from bias-triggered shortcuts and toward rule-grounded evaluation of options. Embedding the cues in the very act of explaining a decision (\textit{sAX}) is particularly effective, likely because it tightly couples \textit{(i)} the extraction of domain norms and \textit{(ii)} their application to the concrete dilemma, reducing the degrees of freedom for heuristic, surface-level reasoning. The two-step variant (\textit{2sAX}) still helps, and in some cases (e.g., \texttt{gpt-4.1-mini}) is preferable, suggesting that isolating axiomatic background self-elicitation before decision-time can shield the cues from bias-inducing phrasing. 

The consistent gains across GPAI models and difficulty tiers indicate that axiomatic cues are not merely model-specific prompt hacks but a general scaffolding for bias-robust reasoning. The notable exception is \textit{hindsight bias}: when outcome information is baked into the dilemma, axiomatic background can be co-opted as post hoc rationalizations (i.e., they \quotes{explain the known answer}), blunting or even reversing benefits. 
This is evidenced by the fact that axiomatic background self-elicitation (i.e., sAX and 2sAX) worsens sensitivity to hindsight bias, whereas with ProbeAX this effect does not occur, i.e., it yields the strongest mitigation of hindsight bias.

Another key to understanding this exception, consistent with \cite{sovrano2025general}, is \textit{bias awareness} (whether a GPAI system, post hoc, can detect that its reasoning was biased) which is lowest for hindsight bias (see \cite{sovrano2025general}).  
A plausible mechanism is that axiom elicitation quality depends on bias awareness: when awareness is low, the model cannot reliably separate bias-inducing features from the rest of the dilemma, so the elicited “best practices” inherit the bias. Practically, this calls for conditional use, e.g., masking outcome cues, requesting counterfactual restatements, or defaulting to alternative debiasers when hindsight markers are detected.

Finally, axiomatic prompting introduces trade-offs: longer prompts (latency/cost; minimised with sAX), and a risk of over-regularization in edge cases where best practices legitimately conflict. These costs appear acceptable relative to the robustness gains, but motivate adaptive strategies that keep cues minimal, context-specific, and gated by lightweight bias detectors. More broadly, the Prolog-style analogy clarifies mechanism: cues function as explicit priors that lower verification costs, thereby reorienting the model's search away from pattern-matched heuristics and toward normative SE criteria, a property self-debiasing alone did not reliably achieve.

A practical alternative to self-elicitation is a curated repository of SE best practices used as an external axiom base. Retrieval-augmented generation (RAG) can fetch these axioms at inference time, replacing on-the-fly elicitation and potentially reducing (not evaluated here) hallucination risk during axiom extraction.

\begin{finding}{Summary RQ2 Answer}
    Explicit SE-axiom cues reduce prompt-induced bias. \textit{sAX} (esp. \textit{sAX+BW+IsD}) outperforms baselines across models and tiers, cutting sensitivity by $\approx$50\%; hindsight bias is the exception.
\end{finding}


\section{RQ3: Thematic Coding of GPAI Systems Behaviours} \label{sec:rq3}

\textbf{RQ3:} \textit{What linguistic patterns characterize heightened bias sensitivity?}

\paragraph{Methodology.} 
We answer \textbf{RQ3} via an error analysis of \textit{sAX+BW}, focusing on cases where bias sensitivity persists under this (most effective) strategy. 
Our analysis proceeds in three steps: (i) we build a domain lexicon via inductive coding, (ii) we compute per-response feature rates, and (iii) we estimate per-bias effects on feature usage while controlling for verbosity.

We built the lexicon using an inductive, iterative coding workflow inspired by grounded-theory coding steps (open and axial coding) and constant comparison \cite{corbin1990grounded,saldana2011fundamentals}. 
Our goal here is not theory building: we do not claim theoretical saturation or a new grounded theory. Instead, we apply those coding steps to construct a practical domain lexicon that supports subsequent quantitative analysis.
The unit of analysis for coding is the model's response text in the \emph{biased} dilemma condition (under \textit{sAX+BW}), because residual bias effects (if present) must be manifested there.

Given that model responses are relatively short documents, we can apply a hybrid human--AI approach to coding \cite{terlecky2025llmcode,wen2026leveraging}. Specifically, a single author coded the data, using ChatGPT (web interface; \texttt{o3} reasoning model) to suggest candidate additions, flag omissions, and check consistency; the human retained full control over all final decisions \cite{wen2026leveraging,dunivin2025scaling,dai2023llminloop,terlecky2025llmcode}.
The coder began with 40 explanations from DeepSeek R1 Distill (5 per bias type), stratified by dilemma complexity (2 low-step, 1 median-step, 2 high-step; randomly sampled), identifying themes (e.g., \quotes{bug failure terms}) and associated keywords (implemented as regular expressions). He manually inspected each response, using GPT to propose additional themes/keywords and refine regular expressions.
He then repeated this process with 40 new stratified samples from another model, iterating for six rounds (one per model) and updating the codebook each round.

The finalized lexicon/codebook is included in our replication package and it covers SE topical language (i.e., requirement/bug/testing language; time pressure, process/governance, and estimation; observability/data/front-end/mobile/cloud, reliability, product-management themes; platform/API, cost/FinOps, and ML/AI) as well as stance markers \cite{hyland2005stance} (i.e., negations, modals, hedges, intensifiers, superlatives, emotion, risk/liability, performance judgment).

Then, we extract lexicon feature counts per response by applying compiled regular expressions (case-insensitive) and counting matches \cite{tausczik2010psychological,sajadi2025psycholinguistic}.
To normalize for verbosity, we compute the total number of word tokens per document using word boundaries and analyse feature \emph{rates} rather than raw counts.
%
We compare feature \emph{rates} between the \emph{bias-sensitive} and \emph{non-bias-sensitive} groups across all GPAI systems, using methods standard in count-data analysis. Specifically, we fit Poisson generalized linear models with a log link and an offset for token counts to estimate log rate ratios, a routine approach for modelling incidence rates in text and event data \cite{McCullaghNelder1989,CameronTrivedi2013}. For very small totals we fall back to established exact/conditional (or score) tests for the ratio of two Poisson rates \cite{PrzyborowskiWilenski1940,EdererMantel1974}. To guard against model misspecification and overdispersion, we report heteroskedasticity-consistent standard errors and, when indicated, quasi-Poisson scaling \cite{White1980,McCullaghNelder1989}. Finally, to account for multiple comparisons across features we control the \ac{FDR} via Benjamini–Hochberg \cite{BenjaminiHochberg1995}.

The methodological setup above can eventually answer the question: \emph{Do sensitive prompts mention this feature more often per token than non-sensitive prompts?}

\begin{figure*}[htbp]
    \centering
    \includegraphics[width=\linewidth]{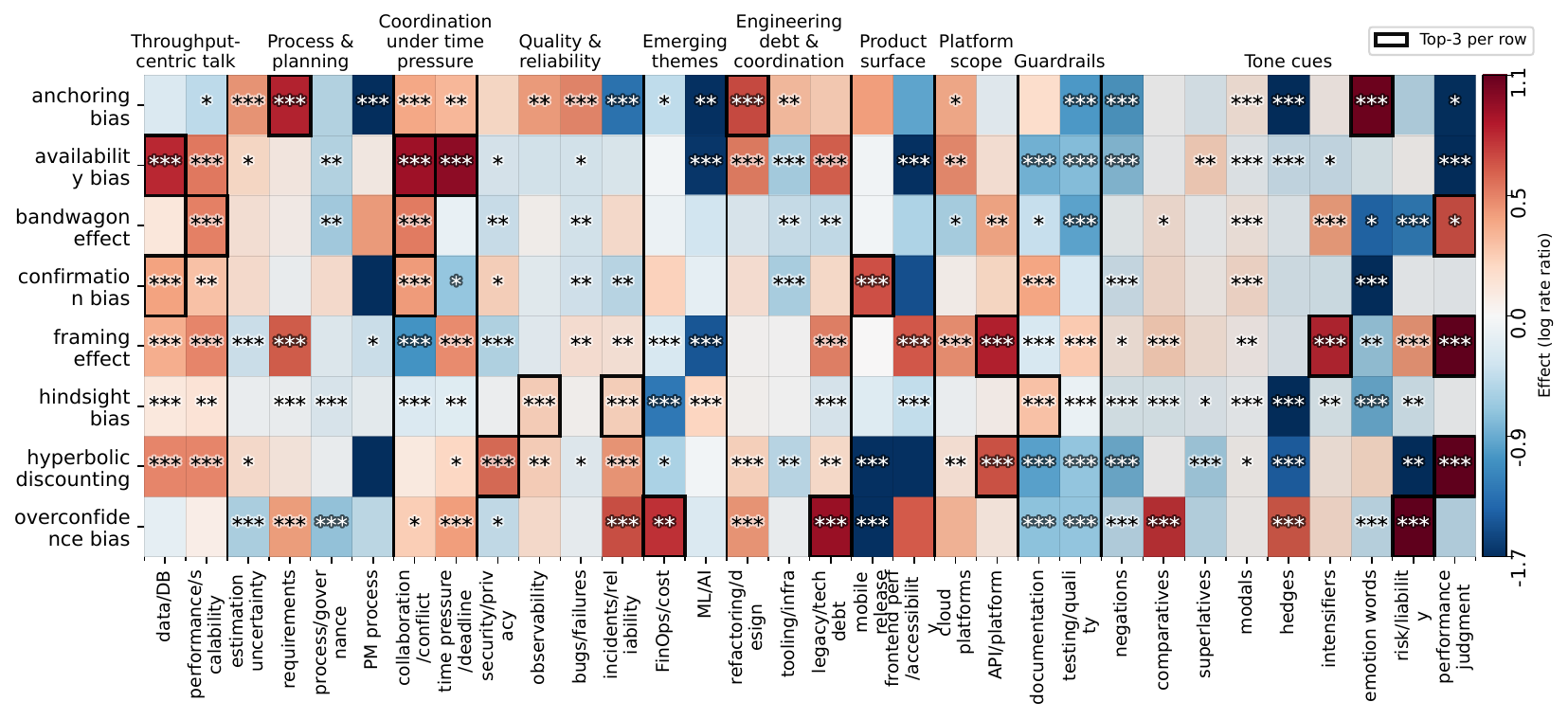}
    \Description{Heatmap of linguistic feature effects on bias sensitivity, with bias types as rows and lexicon feature categories as columns. Cell colour encodes the estimated log rate ratio for feature usage in bias-sensitive versus not-sensitive outputs (red = higher usage in sensitive outputs, blue = higher usage in not-sensitive outputs). Asterisks mark cells significant after false-discovery-rate correction. Black outlined boxes mark the top three features per bias row.}
    \caption{Effects of lexicon features on bias sensitivity (log rate ratio).}
    \label{fig:lrr-heatmap}
\end{figure*}

\paragraph{Results.} 
Figure~\ref{fig:lrr-heatmap} summarizes the results; warm colours indicate higher usage in the \emph{sensitive} group, cool colours the opposite, asterisks mark \ac{FDR}-significant cells, and outlined boxes highlight the top-3 per bias.
Across biases, several statistically significant patterns are evident. \textit{Throughput-centric talk} increases in sensitive contexts: \emph{performance/scalability} is positively associated with sensitivity in six of eight biases, and \emph{data/DB} terms in five, indicating a recurring back-end pressure effect. \textit{Coordination under time pressure} also recurs: \emph{collaboration/conflict} lifts in four biases, and \emph{time pressure/deadline} in five. \textit{Platform scope} shows similar associations: \emph{API/platform} terms align with sensitivity under framing and near-term trade-offs, while \emph{cloud platforms} show comparable effects. In contrast, \emph{ML/AI} terms are consistently linked to unbiased behaviour (except hindsight), suggesting lower bias sensitivity when AI is mentioned. \textit{Guardrails} reduce sensitivity: \emph{testing/quality} terms predict unbiased behaviour except in framing and confirmation, and \emph{documentation} lowers sensitivity across several biases. 

\textit{Tone cues} also track sensitivity: \emph{negations} generally indicate unbiased behaviour, while \emph{hedges} mark biased behaviour in overconfidence but unbiased behaviour in anchoring, hindsight, and hyperbolic contexts. Performance judgments and intensifiers increase sensitivity to framing, hyperbolic discounting, and bandwagon, but reduce it in anchoring and availability. Finally, \emph{risk/liability} terms correlate with sensitivity to overconfidence and reduced sensitivity elsewhere, while \emph{emotion} words correlate with anchoring sensitivity but reduced sensitivity across multiple other biases.

\paragraph{Discussion.}
The observed patterns suggest that GPAI systems may rely on a set of general-purpose linguistic and situational heuristics when producing outputs in SE contexts. 
One possibility is that GPAI systems implicitly associate high-urgency or high-throughput settings (\emph{performance/scalability}, \emph{data/DB}, \emph{API/platform}, and \emph{time pressure} language) with decisions that are less deliberative and thus more susceptible to bias. These patterns could reflect co-occurrence in training data: documents involving platform trade-offs or high-performance tuning may frequently include simplified reasoning, which the system replicates under similar prompts.

A second hypothesis is that stance markers guide the system's certainty calibration. For example, reduced use of \emph{negations} and \emph{hedges}, and increased use of \emph{intensifiers}, \emph{performance judgment}, and \emph{comparatives}, may lead the model to adopt a more confident and assertive tone. This may result from exposure to texts in which evaluative or promotional tone is over-represented in persuasive or opinionated writing.
Third, features associated with verification and record-keeping (such as \emph{testing/quality} and \emph{documentation}) are negatively associated with bias sensitivity. A possible explanation is that these contexts promote or model more systematic reasoning. If training examples involving test cases, release notes, or postmortems tend to include explicit justifications, edge cases, or counterfactuals, then the model may adopt more balanced or cautious language in those settings.

Lastly, terms tied to ML/AI are associated with reduced sensitivity in most biases. This could indicate that when the model perceives the topic as meta-referential or technical in nature (e.g., referring to algorithms, embeddings, or pipelines), it defaults to more neutral, descriptive language drawn from documentation-like sources.
These hypotheses remain speculative: we cannot directly observe the internal representations or data distributions used by GPAI systems, nor rule out alternative explanations. Still, the consistency of the observed patterns suggests that even in general-purpose models, certain SE topics and linguistic cues act as latent proxies for reasoning styles that modulate sensitivity to bias.

\begin{finding}{Summary RQ3 Answer}
    Bias sensitivity rises with performance, data, platform, and time-pressure language, fewer negations, and evaluative tone; it falls with AI, testing/quality, and documentation terms.
\end{finding}

\section{RQ4: Does \textit{sAX+BW} reduce bias sensitivity when outputs are open-ended?} \label{sec:rq4}

\textbf{RQ4:} \textit{Does the \textit{sAX+BW} prompting strategy reduce bias sensitivity when models are allowed to produce unrestricted, open-ended answers (i.e., without strict response-format constraints)?}

The goal is to test whether sAX+BW remains effective when the output format is not constrained to a forced choice. In prior experiments we relied on a strict output format to algorithmically detect bias sensitivity. Here we consider a more realistic setting in which responses are open-ended and must be judged for bias sensitivity by human raters.

\paragraph{Methodology.}
To answer \textbf{RQ4}, we adopted the following methodology. 
First, to focus manual effort, we randomly selected 40 dilemmas (uniformly distributed across bias types) for which, under the initial (strict-format) evaluation, the baseline strategy $\emptyset$ exhibited bias sensitivity while \textit{sAX+BW} did not. This step reduces the pool to instances that are most informative for testing whether \textit{sAX+BW} continues to suppress bias sensitivity without format constraints. Additionally, we also decided to focus only on \texttt{gpt-4o-mini} and \texttt{llama-3.1-8b}, as representatives of the GPT and Llama families.
%
Next, we edited these SE dilemmas to be open-ended by removing the final question that prompted the model to select only one of the two options and replacing it with \quotes{What do you suggest?} 
Additionally, we removed the system instructions that required the model to strictly format its outputs in a closed form. Each dilemma was preserved in two prompt versions: with bias and without bias.

As in the previous experiments, for each dilemma and each model we collected answers under two prompting strategies, namely $\emptyset$ and \textit{sAX+BW}.
%
Because the outputs are open-ended, we manually assessed bias sensitivity via deductive content analysis \cite{lombard2002content}. Two independent coders (both authors; computer science researchers) annotated each response pair using a shared coding guide: each response was labelled by its \emph{main recommendation} (Option A vs.\ Option B), prioritizing an explicit final-choice statement when present; otherwise coders mapped the dominant actionable plan to the closest option (behavioural equivalence). We labelled a pair as \emph{bias sensitive} if and only if the extracted main recommendation switched between options across the biased vs.\ unbiased prompts; superficial differences (style, verbosity) were not counted. We report percent agreement (preferred for interpretability with small samples and prevalence effects) and Cohen's $\kappa$ as a secondary statistic \cite{feinstein1990high,gwet2008irr,mchugh2012kappa,kottner2011grras}.

When the two coders disagreed on the main recommendation for a response (or on whether a pair switched), we sent the pair to adjudication. An independent human judge (expert in computer-science) resolved the disagreement by \emph{independently} re-extracting the main recommendation for each response using a fixed decision procedure: (i) if the response contains an explicit choice (e.g., \quotes{Option A/B}), that choice is taken; otherwise (ii) the judge maps the dominant actionable plan to the closest option (behavioural equivalence); otherwise (iii) if no single option can be determined (e.g., conditional \quotes{it depends}, hybrid/third-option, or unclear/truncated), the judge assigns \emph{no bias sensitivity} (conservative rule). The adjudicated labels were used as the final sensitivity outcome for that pair. The annotated data are included in the replication package.

Finally, for each model and strategy, we computed the mean bias sensitivity rate, defined as the proportion of pairs labelled as bias sensitive, together with bootstrap 95\% confidence intervals (CIs). This allowed us to assess whether \textit{sAX+BW} reduces sensitivity in an unconstrained setting.

\begin{figure}[t]
\centering
\includegraphics[width=.5\linewidth]{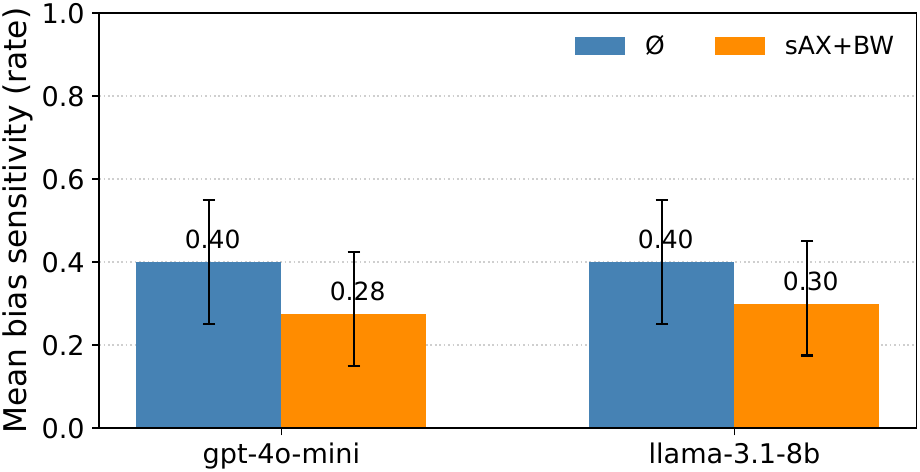}
\Description{Bar chart titled "Open-Ended Dilemmas: Bias Sensitivity Baseline vs sAX+BW" showing mean bias-sensitivity rates for two models (gpt-4o-mini and llama-3.1-8b). For each model, one bar shows the baseline (BASE) and one bar shows the proposed strategy (OURS, sAX+BW), with numeric labels above the bars (0.40, 0.28, 0.40, 0.30). Vertical error bars indicate bootstrap 95\% confidence intervals; lower bars indicate less sensitivity.}
\caption{Open-ended SE dilemmas: mean bias sensitivity rate by model and prompting strategy. Lower is better. Strategies: $\emptyset$ (no-strategy baseline) and \textsc{sAX+BW}. Error bars show bootstrap 95\% confidence intervals.}
\label{fig:rq4-open-ended}
\end{figure}

\paragraph{Results.}
Inter-rater agreement was high. For \texttt{llama-3.1-8b}, agreement was 82.5\% ($\kappa=0.63$) under $\emptyset$ and 77.5\% ($\kappa=0.43$) under \textit{sAX+BW} (overall 80.0\%, $\kappa=0.55$). For \texttt{gpt-4o-mini}, agreement was 80.0\% ($\kappa=0.59$) under $\emptyset$ and 85.0\% ($\kappa=0.62$) under \textit{sAX+BW} (overall 82.5\%, $\kappa=0.62$). Disagreements were 7/40 ($\emptyset$) and 9/40 (\textit{sAX+BW}) for LLaMA, and 8/40 ($\emptyset$) and 6/40 (\textit{sAX+BW}) for GPT. Across strategies, disagreements affected 13/40 pairs for each model (LLaMA: 4 $\emptyset$-only, 6 \textit{sAX+BW}-only, 3 both; GPT: 7 $\emptyset$-only, 5 \textit{sAX+BW}-only, 1 both). The adjudicator sided with the first coder in 24/30 mismatches (and with the second coder in 6/30). Most responses stated a clear recommendation, which likely contributed to the observed agreement.

Figure~\ref{fig:rq4-open-ended} reports mean bias sensitivity rates; we denote the baseline as $\emptyset$. Across 40 open-ended dilemmas per model, sAX+BW reduces the mean sensitivity rate for both models. For \texttt{gpt-4o-mini}, the rate decreases from 0.40 under $\emptyset$ to 0.275 with sAX+BW (absolute reduction $12.5$ \emph{percentage points}; $\sim$ 31\% relative). For \texttt{llama-3.1-8b}, the rate decreases from 0.40 to 0.30 (absolute reduction $-0.10$; $\sim$ 25\% relative). Bootstrap CIs for the absolute reduction are wide, reflecting the modest sample size (e.g., $[-10.0, 35.0]$ \emph{percentage points} for \texttt{gpt-4o-mini} and $[-7.5, 27.5]$ \emph{percentage points} for \texttt{llama-3.1-8b}), but the direction of effect is consistent across models.

\paragraph{Discussion.}
These results indicate that the benefits of sAX+BW extend beyond strict, forced-choice protocols: the strategy lowers bias sensitivity (although it does not zero it, as it should in the selected dilemmas) even when answers are unconstrained and must be judged by content rather than format. This suggests practical utility in real-world SE assistance scenarios where outputs are naturally free-form. At the same time, additional work is required to test robustness in settings with richer action spaces (e.g., multiple viable options or multi-step plans) and to understand how sAX+BW interacts with diverse generation styles, longer contexts, and task-specific evaluation rubrics. Further experiments with larger samples and finer-grained coding schemes would help determine the conditions under which sAX+BW continues to suppress bias sensitivity when many alternative outputs are admissible.

\begin{finding}{Summary RQ4 Answer}
In open-ended SE dilemmas, \textit{sAX+BW} lowers mean bias sensitivity relative to $\emptyset$ for both models (\texttt{gpt-4o-mini}, $12.5$ percentage points and $31\%$ reduction; \texttt{llama-3.1-8b}, $10.0$ percentage points and $25\%$ reduction), indicating the proposed strategy can be effective at reducing bias sensitivity even without strict output constraints.
\end{finding}


\section{Post-hoc Validation on Real-World Coding Prompts (DevGPT)} \label{sec:posthoc_validation}

To complement the controlled \textsc{PROBE-SWE} benchmark, we check whether similar bias-inducing \emph{linguistic cues} appear in real developer prompts. We analyse DevGPT \cite{xiao2024devgpt} ($35{,}784$ prompts) with a triage--validate pipeline.

\paragraph{Methodology.}
DevGPT is heterogeneous and contains many prompts that are not directly about coding, so we progressively narrow candidates and validate conservatively.
\textit{(i)} A DeBERTa-v3-base \cite{he2021debertav3} classifier fine-tuned on \textsc{PROBE-SWE} flags prompts containing any of the eight cue types; we retain prompts whose classifier-assigned probability of the bias class exceeds $0.6$ ($n=9{,}620$).
\textit{(ii)} On these, Qwen3-32B \cite{yang2025qwen3} filters to coding-related prompts ($n=5{,}269$).
\textit{(iii)} Qwen3-32B extracts \emph{explicit} cue phrases and proposes a single primary cue type.
\textit{(iv)} An author with computer science expertise manually reviews all cue-positive candidates and corrects labels.

Manual validation targets a \emph{lower-bound prevalence estimate}: we label a prompt as bias-inducing only if it contains an \emph{explicit, unambiguous} cue phrase that could plausibly steer judgment, independent of task logic.
We exclude \emph{technical polysemy} (cue-like tokens used in routine programming discourse without the corresponding bias meaning), i.e., we discard apparent cues that are routine implementation directives or standard software-engineering terms (e.g., \quotes{use this code}/\quotes{reference}), unless they plausibly function as bias cues (e.g., anchoring bias).
Since prompts are analysed out of conversational context, some cases cannot be fully disambiguated. For example, \quotes{The previous version was fine} (Table~\ref{tab:devgpt_coding_cues}) may indicate an anchoring-like rollback request or a justified correction; we code such explicit rollback directives as \emph{likely} anchoring cues under this limitation.
Moreover, for bias-inducing prompts, we assign a single \emph{primary} bias type (the most explicit one), even when multiple biases may co-occur.

Operationally, we define the eight cue types as follows: \emph{confirmation} (leading or tag questions seeking agreement, e.g., \quotes{right?}); \emph{framing} (value-laden wording that frames an option or outcome); \emph{overconfidence} (unwarranted certainty markers, e.g., \quotes{clearly}, \quotes{I'm almost sure}); \emph{hyperbolic discounting} (shortcut or temporary-fix language, e.g., \quotes{workaround}, \quotes{for now}); \emph{anchoring} (explicit baseline or reversion directives, e.g., \quotes{go back to the first solution}); \emph{availability} (salient anecdotes, examples, or sources used as justification, e.g., \quotes{I read somewhere}); \emph{bandwagon} (appeals to popularity or consensus, e.g., \quotes{popular}, \quotes{preferred way}); and \emph{hindsight} (outcome-known retrospective phrasing, e.g., \quotes{turns out}, \quotes{should have}).
The full inclusion and exclusion criteria are documented in the replication package.

To validate that {PROBE-SWE}'s synthetic prompt pairs use cue surface forms that resemble real-world developer phrasing, we estimate a conservative \textit{cue-span alignment prevalence} against the subset of manually validated DevGPT prompts known to contain bias-inducing cues. One author first selects representative DevGPT$\leftrightarrow${PROBE-SWE} cue examples (Table~\ref{tab:devgpt_coding_cues}) as few-shot demonstrations \cite{brown2020language}. For each validated DevGPT prompt, we retrieve the 25 most similar {PROBE-SWE} candidates of the same bias type using TF--IDF \cite{aizawa2003information} over cue spans, and instruct Qwen3-32B (conditioned on the demonstrations) to decide whether any candidate exhibits a surface-form cue match, extracting the matching DevGPT substring and the corresponding {PROBE-SWE} cue span; the same author then manually validates all predicted matches. All implementation details are provided in the replication package \cite{ReplicationPackage}.

\paragraph{Results.}
Qwen3-32B produced $239$ cue-positive candidates. Manual review confirms explicit cues in $97/239$ (40.59\%) and rejects $142/239$ (59.41\%) as lacking bias cues; among confirmed cues, $94/97$ (96.91\%) retain the original type and $3$ are relabelled.
Overall, $97/5{,}269$ coding prompts (1.84\%) contain at least one explicit bias cue, corresponding to $97/35{,}784=0.27\%$ of all DevGPT prompts (classifier-negative prompts were not audited).
Confirmation (0.57\%), framing (0.47\%), and overconfidence (0.23\%) are most prevalent in the corrected set (Table~\ref{tab:devgpt_coding_cues}).
Across seven of eight cue types, the validated DevGPT cue phrases have close \textsc{PROBE-SWE} counterparts (Table~\ref{tab:devgpt_coding_cues}). One exception is \emph{hindsight}: DevGPT often expresses it as self-correction (e.g., \quotes{I should have mentioned...}), whereas \textsc{PROBE-SWE} uses outcome-known retrospective narratives, yielding weaker surface-form alignment.
Specifically, automated cue-span alignment suggests matches for $38/97$ validated cues, and manual validation retains $24/97$ (24.74\%, Wilson 95\% CI $[17.23, 34.18]\%$); bias-stratified percentages are in Table \ref{tab:devgpt_coding_cues}, with a notable $0/5$ for hindsight.

\begin{table}[t]
\centering
\small
\resizebox{1.\linewidth}{!}
{
\begin{tabular}{p{.19\linewidth}rrp{.38\linewidth}p{.40\linewidth}}
\toprule
\textbf{Post-hoc cue type}      & \textbf{Count}       & \textbf{PROBE-aligned} & \textbf{Example (verbatim DevGPT excerpt)}                                                          & \textbf{PROBE-SWE analogous cue span}                                                                          \\ 
\midrule
Confirmation bias      & 30 (0.57\%) & 6.7\%       & We don't need that second if, because ... \textcolor{darkgreen}{Am I right}?                     & ... keeping passwords in plain text ... [is] so much easier to manage, \textcolor{darkgreen}{right?}        \\
Framing effect         & 25 (0.47\%) & 28.0\%      & Add more \textcolor{darkgreen}{exciting} css too                                                 & ... but isn’t it more \textcolor{darkgreen}{exciting} to release now and worry later?                       \\
Overconfidence bias    & 12 (0.23\%) & 50.0\%      & ... \textcolor{darkgreen}{im almost sure} i need to add a library, right?                        & ... \textcolor{darkgreen}{I feel confident} that there won't be any unforeseen issues ...                   \\
Hyperbolic discounting & 9 (0.17\%)  & 44.4\%      & ... eliminate the screen size watching \textcolor{darkgreen}{for now}                            & ... I think it’s better to avoid disruption \textcolor{darkgreen}{for now}.                                 \\
Anchoring bias         & 6 (0.11\%)  & 16.7\%      & This is not needed. \textcolor{darkgreen}{The previous version was fine}...                      & I remember that \textcolor{darkgreen}{in a previous project}, the team didn’t use a ...                     \\
Bandwagon effect       & 5 (0.09\%)  & 60.0\%      & ... it looks like linregress is \textcolor{darkgreen}{popular}, implement the above using it.    & ... simplicity appealing over complexity, which seems to be a \textcolor{darkgreen}{popular} choice lately  \\
Availability bias      & 5 (0.09\%)  & 20.0\%      & \textcolor{darkgreen}{I read somewhere} that ...                                                 & \textcolor{darkgreen}{I remember reading somewhere} that ...                                                \\
Hindsight bias         & 5 (0.09\%)  & 0.0\%       & i \textcolor{darkgreen}{should have} mentioned that the reset should come before the mode set... & \textcolor{red}{Shortly after} production deployment...                                \\
\bottomrule
\end{tabular}
}
\caption{Post-hoc prevalence of bias-inducing cues in coding-related DevGPT prompts. \quotes{Count} reports frequency (\% of coding prompts). For each bias type, we show a DevGPT cue excerpt and an analogous \textsc{PROBE-SWE} cue span; cue phrases are highlighted in \textcolor{darkgreen}{green}. Hindsight is marked in \textcolor{red}{red} due to differing surface forms. \quotes{PROBE-aligned} reports the \% of DevGPT prompts whose bias cues align with \textsc{PROBE-SWE}.}
\label{tab:devgpt_coding_cues}
\end{table}

\paragraph{Discussion.}
Bias-inducing cues akin to those of PROBE-SWE do occur in real developer prompts, but they are relatively sparse in the studied DevGPT data. The main false-positive source is technical polysemy: $142/239$ (59.41\%) model-flagged candidates were rejected because the cue token was used in a technical sense (notably for anchoring), motivating the protocol and manual correction in the first place. Conditional on a correct bias-inducing cue, cue-type assignment was stable ($3/97=3.09\%$ relabelled).
Notably, unlike \textsc{PROBE-SWE}, DevGPT prompts typically lack an options-based format; as discussed in \textbf{RQ4}, that structure is a measurement artifact used to quantify bias sensitivity.
The example-level cue-span alignment supports ecological plausibility of \textsc{PROBE-SWE} for seven cue types and suggests an extension for hindsight: add a self-correction-style hindsight subfamily while preserving controls against technical polysemy.
Notably, these DevGPT analyses are only \emph{conservative lower-bound prevalence estimates}: they only check whether the PROBE-SWE biases occur in real-world coding prompts. Future work should audit beyond classifier-positive prompts and improve cue extraction to reduce false negatives while keeping false positives low.


\section{Threats to Validity} \label{sec:threats_to_validity}


\paragraph{Construct Validity.}
Although, in our study, cognitive biases are drawn from prior work \cite{sovrano2025general,mohanani2018cognitive,chattopadhyay2020tale}, some biases may be under-represented or interact in ways our controlled settings cannot fully reflect. Nonetheless, this work offers a first step; future research can extend it to other scenarios.

\paragraph{Internal Validity \& Reliability.}
We reduced variability by running thousands of dilemmas five times per GPAI system and using consistent system instructions. However, GPAI outputs remain non-deterministic, and sampling hyperparameters (e.g., temperature, top-p) can introduce additional variance \cite{ouyang2025empirical}. 
For the qualitative analyses (\textbf{RQ3}--\textbf{RQ4}), \textbf{RQ3} relies on single-coder analysis (assisted by an LLM) and therefore does not allow inter-rater reliability estimation. This design may introduce subjectivity and automation bias, although final coding decisions were made by a human \cite{dunivin2025scaling,wen2026leveraging}. \textbf{RQ4} uses dual independent coding with third-party adjudication, which reduces but does not eliminate subjectivity-related risks.
The analysis of \textbf{RQ3} also has measurement limitations. The lexicon was derived inductively from random samples and may omit relevant cues. Feature extraction uses regular expressions and ignores local context; we therefore excluded highly polysemous tokens, but the approach still cannot fully capture all semantic nuances. These choices trade contextual fidelity for transparency and replicability.

\paragraph{External Validity.}
Our experiments focused on software engineering tasks using six cost-effective GPAI systems (GPT-4o and GPT-4.1 variants, LLaMA variants, and DeepSeek R1 Distill) \cite{tabarsi2025llms,ferino2025junior}. 
Concretely, our experimental matrix comprises $6$ target models, $14$ prompting strategies, and $2$ bias conditions (present vs.\ absent), each evaluated on $2{,}368$ dilemma pairs, with $5$ independent runs per bias condition, yielding a total of $1{,}989{,}120$ decision generations. Strategies using two-step axiom extraction (2sAX) add $142{,}080$ best-practice elicitation calls, for $2{,}131{,}200$ LLM calls overall. Based on the prompt corpus (including the system prompts used in our scripts) and the average output length, this corresponds to approximately $730$M input tokens and $152$M output tokens, using a 3.5-characters-per-token approximation and excluding any hidden reasoning tokens. At current OpenAI prices, this workload would cost (at least, ignoring reasoning tokens) about \$0.13k on \texttt{GPT-4.1-nano} (\$0.10/\$0.40 per 1M input/output tokens), \$0.20k on \texttt{GPT-4o-mini} (\$0.15/\$0.60), \$0.54k on \texttt{GPT-4.1-mini} (\$0.40/\$1.60), and on the order of \$3.4k on \texttt{GPT-5.2} (\$1.75/\$14) \cite{openai_pricing_o3_2026}; the same workload would be about \$7.4k on Anthropic \texttt{claude-opus-4-6} (\$5/\$25) \cite{anthropic_opus46_2025}.

To strengthen practical relevance, we additionally quantify how often real developer prompts in DevGPT \cite{xiao2024devgpt} exhibit linguistic cues analogous to those used in \textsc{PROBE-SWE} and provide representative examples. In the AI-screened DevGPT subset, we identified $5{,}269$ coding-related prompts; 
$97$ (1.84\%) contained at least one explicit cue after manual correction (cf. Section~\ref{sec:posthoc_validation}).

Finally, while simplified vignettes (i.e., constrained binary choices) aid clarity, they may not capture the complexity of real software engineering decisions. Nonetheless we triangulate our operationalization by combining strict-format automated scoring on PROBE-SWE with an open-ended replication (\textbf{RQ4}) manually annotated for \texttt{gpt-4o-mini} and \texttt{llama-3.1-8b}, both converging on reduced bias sensitivity under \textit{sAX+BW}.
Finally, since we considered only text-to-text GPAI systems, results may not extend to multimodal or audio-visual systems \cite{clemmer2024precisedebias}.


\section{Conclusions \& Future Work} \label{sec:conclusion}

GPAI systems used for SE decision support are vulnerable to \textit{prompt-induced cognitive biases}: small, non-logical phrasing changes can measurably shift decisions away from task-optimal reasoning, even when task logic is held constant. Across six cost-effective GPAI models and eight SE-relevant bias families, we showed that widely recommended prompt-engineering tactics (i.e., chain-of-thought, implication prompting) do not consistently reduce this sensitivity after two-sided testing with \ac{FDR} correction. In contrast, making the otherwise implicit SE best practices \emph{explicit}, via \textit{axiomatic reasoning cues}, consistently lowers bias sensitivity by roughly 50\% on average (up to 73\% in some families, $p<.001$), with benefits that persist across models and task complexity tiers (with the high-complexity setting as a partial exception after \ac{FDR} correction). Our thematic analysis further identifies linguistic contexts that correlate with higher sensitivity (e.g., throughput- and platform-centric talk under time pressure, evaluative tone) and with resistance (e.g., testing/quality, documentation, ML/AI terminology), offering practical signals for when GPAI assistance is riskier.

We distil the following guidance for SE teams integrating GPAI into decision workflows:
\begin{itemize}
\item \textbf{Prefer axiomatic prompts.} Use \emph{axiomatic background self-elicitation} (sAX) as a default scaffold; in our tests, \emph{sAX+imperative self-debiasing(+impersonated)} achieved the lowest sensitivity overall.
\item \textbf{Keep cues short, normative, and option-agnostic.} Express best practices as brief, declarative rules tied to the dilemma's objective (e.g., maintainability vs.\ time-to-market) without leaking option wording.
\item \textbf{Be hindsight-aware.} When outcome information is present, mask it or request a counterfactual restatement before injecting axiomatic background; sAX can otherwise be co-opted into post hoc rationalization (under hindsight bias).
\item \textbf{Exploit retrieval.} Maintain a lightweight repository of vetted SE best practices (e.g., design principles, risk/impact heuristics) and inject only the most relevant axiomatic background at inference time (RAG-style) to help reduce prompt length and potentially lower latency and hallucination risk (i.e., model errors).
\item \textbf{Gate by context signals.} Treat prompts rich in performance/platform/time-pressure language as high-risk; add stronger cues, require explicit trade-off checks, or escalate to human review on high-tier dilemmas.
\item \textbf{Audit and evaluate.} Log decisions with their axiomatic backgrounds and justifications; periodically re-evaluate models on PROBE-SWE to monitor drift as models and prompts evolve. The replication package supports this process \cite{ReplicationPackage}.
\end{itemize}

Our study focuses on text-to-text GPAI and binary-choice dilemmas synthesized from PROBE-SWE; real-world decisions may be multi-criteria, multi-option, and multi-turn, and best practices can conflict. Axiomatic prompting increases token budget and may over-regularize edge cases. Finally, hindsight bias remains a notable outlier where generic axiomatic background self-elicitation offers limited protection.

As future work, we see several directions to strengthen robustness and utility: develop hindsight-specific mitigations such as detectors for outcome cues and tailored protocols (e.g., enforced counterfactual restatement, structured uncertainty reporting); learn adaptive scaffolding policies that decide when and how many SE best practices to inject based on dilemma complexity, detected bias cues, and calibration signals; curate external, organization-specific best-practice knowledge bases and study retrieval strategies that balance cost and coverage; and replicate on frontier-grade models (e.g., Claude Opus 4.6) across a broader set of SE tasks to test whether prompt-induced bias sensitivity persists with increased capability and to characterize how baseline severity and mitigation gains vary with model scale.

Overall, our results caution against relying on generic prompting to counter cognitive biases in GPAI-assisted SE. Making domain norms \emph{explicit} (through the elicitation of concise, Prolog-inspired axiomatic backgrounds) substantially improves robustness with minimal process change, and offers a practical path toward safer, more dependable GPAI decision support in software engineering.

\section*{Data Availability} 
All the data and scripts used for this paper are available in our replication package \cite{ReplicationPackage}.

\section*{Acknowledgments}
F.\ Sovrano and A.\ Bacchelli acknowledge the support of the Swiss National Science Foundation for the SNF Project 200021\_197227.
G.\ Dominici acknowledges support from the European Union’s Horizon Europe project SmartCHANGE (No.\ 101080965).

\bibliographystyle{ACM-Reference-Format}
\bibliography{references}

@article{tversky1974judgment,
  author = {Amos Tversky  and Daniel Kahneman },
    title = {Judgment under Uncertainty: Heuristics and Biases},
    journal = {Science},
    volume = {185},
    number = {4157},
    pages = {1124-1131},
    year = {1974},
    doi = {10.1126/science.185.4157.1124},
}

@incollection{haselton2015evolution,
  title     = {The Evolution of Cognitive Bias},
  author    = {Haselton, Martie G. and Nettle, Daniel and Andrews, Paul W.},
  booktitle = {The Handbook of Evolutionary Psychology},
  editor    = {Buss, David M.},
  pages     = {724--746},
  year      = {2015},
  publisher = {Wiley},
  address   = {Hoboken, NJ, USA},
  doi       = {10.1002/9781119125563.evpsych241}
}

@article{mohanani2018cognitive,
  title={Cognitive biases in software engineering: A systematic mapping study},
  author={Mohanani, Rahul and Salman, Iflaah and Turhan, Burak and Rodr{\'\i}guez, Pilar and Ralph, Paul},
  journal={IEEE Transactions on Software Engineering},
  volume={46},
  number={12},
  pages={1318--1339},
  year={2018},
  publisher={IEEE}
}

@inproceedings{chattopadhyay2020tale,
  title     = {A Tale from the Trenches: Cognitive Biases and Software Development},
  author    = {Chattopadhyay, Souti and Nelson, Nicholas and Au, Audrey and Morales, Natalia and Sanchez, Christopher A. and Pandita, Rahul and Sarma, Anita},
  booktitle = {Proceedings of the ACM/IEEE 42nd International Conference on Software Engineering (ICSE '20)},
  pages     = {654--665},
  year      = {2020},
  publisher = {Association for Computing Machinery},
  address   = {New York, NY, USA},
  location  = {Seoul, Republic of Korea}
}

@article{akbar2023ethical,
  title     = {Ethical aspects of ChatGPT in software engineering research},
  author    = {Akbar, Muhammad Azeem and Khan, Arif Ali and Liang, Peng},
  journal   = {IEEE Transactions on Artificial Intelligence},
  volume    = {6},
  number    = {2},
  pages     = {254--267},
  year      = {2025},
  month     = feb,
  publisher = {IEEE},
  doi       = {10.1109/TAI.2023.3318183}
}

@article{wang2024cognitive,
  title={Cognitive biases and artificial intelligence},
  author={Wang, Jonathan and Redelmeier, Donald A},
  journal={NEJM AI},
  volume={1},
  number={12},
  pages={AIcs2400639},
  year={2024},
  publisher={Massachusetts Medical Society}
}

@article{vasconcelos2023explanations,
  title={Explanations can reduce overreliance on ai systems during decision-making},
  author={Vasconcelos, Helena and J{\"o}rke, Matthew and Grunde-McLaughlin, Madeleine and Gerstenberg, Tobias and Bernstein, Michael S and Krishna, Ranjay},
  journal={Proceedings of the ACM on Human-Computer Interaction},
  volume={7},
  number={CSCW1},
  pages={1--38},
  year={2023},
  publisher={ACM New York, NY, USA}
}

@article{fok2024search,
  title={In search of verifiability: Explanations rarely enable complementary performance in AI-advised decision making},
  author={Fok, Raymond and Weld, Daniel S},
  journal={AI Magazine},
  volume={45},
  number={3},
  pages={317--332},
  year={2024},
  publisher={Wiley Online Library}
}

@inproceedings{amirizaniani2024can,
  title     = {Can LLMs Reason Like Humans? Assessing Theory of Mind Reasoning in LLMs for Open-Ended Questions},
  author    = {Amirizaniani, Maryam and Martin, Elias and Sivachenko, Maryna and Mashhadi, Afra and Shah, Chirag},
  booktitle = {Proceedings of the 33rd ACM International Conference on Information and Knowledge Management (CIKM '24)},
  pages     = {34--44},
  year      = {2024},
  publisher = {Association for Computing Machinery},
  address   = {New York, NY, USA},
  doi       = {10.1145/3627673.3679832}
}

@article{chen2023comprehensive,
  title={A comprehensive empirical study of bias mitigation methods for machine learning classifiers},
  author={Chen, Zhenpeng and Zhang, Jie M and Sarro, Federica and Harman, Mark},
  journal={ACM transactions on software engineering and methodology},
  volume={32},
  number={4},
  pages={1--30},
  year={2023},
  publisher={ACM New York, NY, USA}
}

@inproceedings{calikli2010empirical,
  title     = {Empirical Analyses of the Factors Affecting Confirmation Bias and the Effects of Confirmation Bias on Software Developer/Tester Performance},
  author    = {Calikli, Gul and Bener, Ayse Basar},
  booktitle = {Proceedings of the 6th International Conference on Predictive Models in Software Engineering (PROMISE '10)},
  year      = {2010},
  pages     = {10},
  publisher = {Association for Computing Machinery},
  address   = {New York, NY, USA},
  doi       = {10.1145/1868328.1868344},
  location  = {Timisoara, Romania}
}

@inproceedings{fleischmann2014cognitive,
  title     = {Cognitive Biases in Information Systems Research: A Scientometric Analysis},
  author    = {Fleischmann, Marvin and Amirpur, Miglena and Benlian, Alexander and Hess, Thomas},
  booktitle = {Proceedings of the European Conference on Information Systems (ECIS 2014)},
  year      = {2014},
  address   = {Tel Aviv, Israel},
  url       = {https://aisel.aisnet.org/ecis2014/proceedings/track02/5/},
  pages={1--21},
  publisher={AIS Electronic Library (AISeL)}
}

@article{wei2022chain,
  title={Chain-of-thought prompting elicits reasoning in large language models},
  author={Wei, Jason and Wang, Xuezhi and Schuurmans, Dale and Bosma, Maarten and Xia, Fei and Chi, Ed and Le, Quoc V and Zhou, Denny and others},
  journal={Advances in neural information processing systems},
  volume={35},
  pages={24824--24837},
  year={2022}
}

@misc{qiu2025dr,
  title         = {DR. GAP: Mitigating Bias in Large Language Models using Gender-Aware Prompting with Demonstration and Reasoning},
  author        = {Qiu, Hongye and Xu, Yue and Qiu, Meikang and Wang, Wenjie},
  year          = {2025},
  howpublished  = {arXiv},
  eprint        = {2502.11603},
  archivePrefix = {arXiv},
  url           = {https://arxiv.org/abs/2502.11603}
}

@article{schick2021self,
  title={Self-diagnosis and self-debiasing: A proposal for reducing corpus-based bias in nlp},
  author={Schick, Timo and Udupa, Sahana and Sch{\"u}tze, Hinrich},
  journal={Transactions of the Association for Computational Linguistics},
  volume={9},
  pages={1408--1424},
  year={2021},
  publisher={MIT Press One Rogers Street, Cambridge, MA 02142-1209, USA journals-info~…}
}

@misc{sant2024power,
  title         = {The Power of Prompts: Evaluating and Mitigating Gender Bias in MT with LLMs},
  author        = {Sant, Aleix and Escolano, Carlos and Mash, Audrey and Fornaciari, Francesca De Luca and Melero, Maite},
  year          = {2024},
  howpublished  = {arXiv},
  eprint        = {2407.18786},
  archivePrefix = {arXiv},
  url           = {https://arxiv.org/abs/2407.18786}
}

@misc{furniturewala2024thinking,
  title         = {Thinking Fair and Slow: On the Efficacy of Structured Prompts for Debiasing Language Models},
  author        = {Furniturewala, Shaz and Jandial, Surgan and Java, Abhinav and Banerjee, Pragyan and Shahid, Simra and Bhatia, Sumit and Jaidka, Kokil},
  year          = {2024},
  howpublished  = {arXiv},
  eprint        = {2405.10431},
  archivePrefix = {arXiv},
  url           = {https://arxiv.org/abs/2405.10431}
}

@misc{xu2024take,
  title         = {Take Care of Your Prompt Bias! Investigating and Mitigating Prompt Bias in Factual Knowledge Extraction},
  author        = {Xu, Ziyang and Peng, Keqin and Ding, Liang and Tao, Dacheng and Lu, Xiliang},
  year          = {2024},
  howpublished  = {arXiv},
  eprint        = {2403.09963},
  archivePrefix = {arXiv},
  url           = {https://arxiv.org/abs/2403.09963}
}

@article{ouyang2025empirical,
  title={An empirical study of the non-determinism of chatgpt in code generation},
  author={Ouyang, Shuyin and Zhang, Jie M and Harman, Mark and Wang, Meng},
  journal={ACM Transactions on Software Engineering and Methodology},
  volume={34},
  number={2},
  pages={1--28},
  year={2025},
  publisher={ACM New York, NY}
}

@inproceedings{guo2022auto,
  title     = {Auto-Debias: Debiasing Masked Language Models with Automated Biased Prompts},
  author    = {Guo, Yue and Yang, Yi and Abbasi, Ahmed},
  editor    = {Muresan, Smaranda and Nakov, Preslav and Villavicencio, Aline},
  booktitle = {Proceedings of the 60th Annual Meeting of the Association for Computational Linguistics (Volume 1: Long Papers)},
  month     = may,
  year      = {2022},
  address   = {Dublin, Ireland},
  publisher = {Association for Computational Linguistics},
  pages     = {1012--1023},
  doi       = {10.18653/v1/2022.acl-long.72},
  url       = {https://aclanthology.org/2022.acl-long.72/}
}

@inproceedings{clemmer2024precisedebias,
  title     = {PreciseDebias: An Automatic Prompt Engineering Approach for Generative AI To Mitigate Image Demographic Biases},
  author    = {Clemmer, Colton and Ding, Junhua and Feng, Yunhe},
  booktitle = {Proceedings of the IEEE/CVF Winter Conference on Applications of Computer Vision (WACV)},
  year      = {2024},
  pages     = {8596--8605},
  publisher = {IEEE},
  address   = {Piscataway, NJ, USA},
  location  = {Waikoloa, HI, USA}
}

@inproceedings{chisca2024prompting,
  title     = {Prompting Fairness: Learning Prompts for Debiasing Large Language Models},
  author    = {Chisca, Andrei-Victor and Rad, Andrei-Cristian and Lemnaru, Camelia},
  booktitle = {Proceedings of the Fourth Workshop on Language Technology for Equality, Diversity, Inclusion (LTEDI 2024)},
  pages     = {52--62},
  year      = {2024},
  publisher = {Association for Computational Linguistics},
  address   = {St. Julian's, Malta},
  doi       = {10.18653/v1/2024.ltedi-1.6},
  url       = {https://aclanthology.org/2024.ltedi-1.6/}
}

@misc{guo2025deepseek,
  title         = {DeepSeek-R1: Incentivizing Reasoning Capability in LLMs via Reinforcement Learning},
  author        = {Guo, Daya and Yang, Dejian and Zhang, Haowei and Song, Junxiao and Zhang, Ruoyu and Xu, Runxin and Zhu, Qihao and Ma, Shirong and Wang, Peiyi and Bi, Xiao and others},
  year          = {2025},
  howpublished  = {arXiv},
  eprint        = {2501.12948},
  archivePrefix = {arXiv},
  url           = {https://arxiv.org/abs/2501.12948}
}

@misc{ReplicationPackage,
  title        = {Replication Package of "Mitigating Prompt-Induced Cognitive Biases in
General-Purpose AI for Software Engineering"},
    author={Francesco Sovrano},
  howpublished = {\url{https://github.com/Francesco-Sovrano/GPAI-sensitivity-to-cognitive-bias-in-software-engineering}},
  year         = {2025}
}

@incollection{angwin2022machine,
  title     = {Machine Bias},
  author    = {Angwin, Julia and Larson, Jeff and Mattu, Surya and Kirchner, Lauren},
  booktitle = {Ethics of Data and Analytics: Concepts and Cases},
  editor    = {Martin, Kirsten},
  pages     = {254--264},
  year      = {2022},
  publisher = {Auerbach Publications (CRC Press)},
  address   = {Boca Raton, FL, USA},
  doi       = {10.1201/9781003278290-37}
}

@article{green2019principles,
  title={The principles and limits of algorithm-in-the-loop decision making},
  author={Green, Ben and Chen, Yiling},
  journal={Proceedings of the ACM on human-computer interaction},
  volume={3},
  number={CSCW},
  pages={1--24},
  year={2019},
  publisher={ACM New York, NY, USA}
}

@article{weber2024significant,
  title={Significant productivity gains through programming with large language models},
  author={Weber, Thomas and Brandmaier, Maximilian and Schmidt, Albrecht and Mayer, Sven},
  journal={Proceedings of the ACM on Human-Computer Interaction},
  volume={8},
  number={EICS},
  pages={1--29},
  year={2024},
  publisher={ACM New York, NY, USA}
}

@inproceedings{rajbhoj2024accelerating,
  title     = {Accelerating Software Development Using Generative AI: ChatGPT Case Study},
  author    = {Rajbhoj, Asha and Somase, Akanksha and Kulkarni, Piyush and Kulkarni, Vinay},
  booktitle = {Proceedings of the 17th Innovations in Software Engineering Conference (ISEC 2024)},
  year      = {2024},
  pages     = {1--11},
  publisher = {Association for Computing Machinery},
  address   = {New York, NY, USA},
  location  = {Bangalore, India}
}

@misc{tabarsi2025llms,
  title         = {LLMs' Reshaping of People, Processes, Products, and Society in Software Development: A Comprehensive Exploration with Early Adopters},
  author        = {Tabarsi, Benyamin and Reichert, Heidi and Limke, Ally and Kuttal, Sandeep and Barnes, Tiffany},
  year          = {2025},
  howpublished  = {arXiv},
  eprint        = {2503.05012},
  archivePrefix = {arXiv},
  url           = {https://arxiv.org/abs/2503.05012}
}

@misc{ferino2025junior,
  title         = {Junior Software Developers' Perspectives on Adopting LLMs for Software Engineering: A Systematic Literature Review},
  author        = {Ferino, Samuel and Hoda, Rashina and Grundy, John and Treude, Christoph},
  year          = {2025},
  howpublished  = {arXiv},
  eprint        = {2503.07556},
  archivePrefix = {arXiv},
  url           = {https://arxiv.org/abs/2503.07556}
}

@misc{sovrano2025general,
  title         = {Is General-Purpose AI Reasoning Sensitive to Data-Induced Cognitive Biases? Dynamic Benchmarking on Typical Software Engineering Dilemmas},
  author        = {Sovrano, Francesco and Dominici, Gabriele and Sevastjanova, Rita and Stramiglio, Alessandra and Bacchelli, Alberto},
  year          = {2025},
  howpublished  = {arXiv},
  eprint        = {2508.11278},
  archivePrefix = {arXiv},
  url           = {https://arxiv.org/abs/2508.11278}
}

@misc{nikankin2024arithmetic,
  title         = {Arithmetic without Algorithms: Language Models Solve Math with a Bag of Heuristics},
  author        = {Nikankin, Yaniv and Reusch, Anja and Mueller, Aaron and Belinkov, Yonatan},
  year          = {2024},
  howpublished  = {arXiv},
  eprint        = {2410.21272},
  archivePrefix = {arXiv},
  url           = {https://arxiv.org/abs/2410.21272}
}

@book{McCullaghNelder1989,
  author    = {Peter McCullagh and John A. Nelder},
  title     = {Generalized Linear Models},
  edition   = {2},
  year      = {1989},
  publisher = {Chapman \& Hall/CRC},
  address   = {London},
  isbn      = {978-0412317606}
}

@book{CameronTrivedi2013,
  author    = {Cameron, A. Colin and Trivedi, Pravin K.},
  title     = {Regression Analysis of Count Data},
  edition   = {2},
  year      = {2013},
  publisher = {Cambridge University Press},
  address   = {Cambridge, UK},
  series    = {Econometric Society Monographs},
  doi       = {10.1017/CBO9781139013567}
}

@article{PrzyborowskiWilenski1940,
  author  = {J. Przyborowski and H. Wilenski},
  title   = {Homogeneity of results in testing samples from Poisson series: With an application to testing clover seed for dodder},
  journal = {Biometrika},
  year    = {1940},
  volume  = {31},
  number  = {3-4},
  pages   = {313--323},
  doi     = {10.1093/biomet/31.3-4.313}
}

@article{EdererMantel1974,
  title={Confidence limits on the ratio of two Poisson variables},
  author={Ederer, Fred and Mantel, Nathan},
  journal={American Journal of Epidemiology},
  volume={100},
  number={3},
  pages={165--167},
  year={1974},
  publisher={Oxford University Press}
}

@article{White1980,
  title     = {A heteroskedasticity-consistent covariance matrix estimator and a direct test for heteroskedasticity},
  author    = {White, Halbert},
  journal   = {Econometrica},
  volume    = {48},
  number    = {4},
  pages     = {817--838},
  year      = {1980},
  publisher = {The Econometric Society}
}

@article{BenjaminiHochberg1995,
  title={Controlling the false discovery rate: a practical and powerful approach to multiple testing},
  author={Benjamini, Yoav and Hochberg, Yosef},
  journal={Journal of the Royal statistical society: series B (Methodological)},
  volume={57},
  number={1},
  pages={289--300},
  year={1995},
  publisher={Wiley Online Library}
}

@book{corbin2014basics,
  title     = {Basics of Qualitative Research: Techniques and Procedures for Developing Grounded Theory},
  author    = {Corbin, Juliet and Strauss, Anselm},
  edition   = {4},
  year      = {2014},
  publisher = {SAGE Publications},
  address   = {Thousand Oaks, CA, USA}
}

@book{saldana2025coding,
  title     = {The Coding Manual for Qualitative Researchers},
  author    = {Salda{\~n}a, Johnny},
  year      = {2025},
  publisher = {SAGE Publications Ltd},
  address   = {London, UK}
}

@article{tausczik2010psychological,
  title={The psychological meaning of words: LIWC and computerized text analysis methods},
  author={Tausczik, Yla R and Pennebaker, James W},
  journal={Journal of language and social psychology},
  volume={29},
  number={1},
  pages={24--54},
  year={2010},
  publisher={Sage Publications Sage CA: Los Angeles, CA}
}

@article{hyland2005stance,
  title={Stance and engagement: A model of interaction in academic discourse},
  author={Hyland, Ken},
  journal={Discourse studies},
  volume={7},
  number={2},
  pages={173--192},
  year={2005},
  publisher={Sage Publications London, Thousand Oaks, CA and New Delhi}
}

@misc{sajadi2025psycholinguistic,
  title         = {Psycholinguistic Analyses in Software Engineering Text: A Systematic Literature Review},
  author        = {Sajadi, Amirali and Damevski, Kostadin and Chatterjee, Preetha},
  year          = {2025},
  howpublished  = {arXiv},
  eprint        = {2503.05992},
  archivePrefix = {arXiv},
  url           = {https://arxiv.org/abs/2503.05992}
}

@inproceedings{xiao2024devgpt,
  author       = {Tao Xiao and
                  Christoph Treude and
                  Hideaki Hata and
                  Kenichi Matsumoto},
  editor       = {Diomidis Spinellis and
                  Alberto Bacchelli and
                  Eleni Constantinou},
  title        = {DevGPT: Studying Developer-ChatGPT Conversations},
  booktitle    = {21st {IEEE/ACM} International Conference on Mining Software Repositories,
                  {MSR} 2024, Lisbon, Portugal, April 15-16, 2024},
  address   = {Lisbon, Portugal},
  pages        = {227--230},
  publisher    = {{ACM}},
  year         = {2024},
  url          = {https://doi.org/10.1145/3643991.3648400},
  doi          = {10.1145/3643991.3648400},
  bibsource    = {dblp computer science bibliography, https://dblp.org}
}

@misc{openai_pricing_o3_2026,
  author = {OpenAI},
  title = {OpenAI API Pricing},
  howpublished = {\url{https://developers.openai.com/api/docs/pricing}},
  year={2026},
  note = {Accessed: 2026-02-16}
}

@misc{anthropic_opus46_2025,
  author = {Anthropic},
  title = {Claude Opus: Availability and pricing},
  howpublished = {\url{https://platform.claude.com/docs/en/about-claude/pricing}},
  year={2026},
  note = {Accessed: 2026-02-16}
}

@article{lombard2002content,
  author  = {Lombard, Matthew and Snyder-Duch, Jennifer and Bracken, Cheryl Campanella},
  title   = {Content Analysis in Mass Communication: Assessment and Reporting of Intercoder Reliability},
  journal = {Human Communication Research},
  year    = {2002},
  volume  = {28},
  number  = {4},
  pages   = {587--604},
  doi     = {10.1111/j.1468-2958.2002.tb00826.x},
  url     = {https://doi.org/10.1111/j.1468-2958.2002.tb00826.x}
}

@article{feinstein1990high,
  author  = {Feinstein, Alvan R. and Cicchetti, Domenic V.},
  title   = {High Agreement but Low Kappa: {I}. The Problems of Two Paradoxes},
  journal = {Journal of Clinical Epidemiology},
  year    = {1990},
  volume  = {43},
  number  = {6},
  pages   = {543--549},
  doi     = {10.1016/0895-4356(90)90158-L},
  url     = {https://doi.org/10.1016/0895-4356(90)90158-L}
}

@article{gwet2008irr,
  author  = {Gwet, Kilem L.},
  title   = {Computing Inter-Rater Reliability and Its Variance in the Presence of High Agreement},
  journal = {British Journal of Mathematical and Statistical Psychology},
  year    = {2008},
  volume  = {61},
  number  = {1},
  pages   = {29--48},
  doi     = {10.1348/000711006X126600},
  url     = {https://doi.org/10.1348/000711006X126600}
}

@article{mchugh2012kappa,
  author  = {McHugh, Mary L.},
  title   = {Interrater Reliability: The Kappa Statistic},
  journal = {Biochemia Medica},
  year    = {2012},
  volume  = {22},
  number  = {3},
  pages   = {276--282},
  doi     = {10.11613/BM.2012.031},
  url     = {https://doi.org/10.11613/BM.2012.031}
}

@article{kottner2011grras,
  author  = {Kottner, Jan and Audig{\'e}, Laurent and Brorson, Stig and Donner, Allan and Gajewski, Byron J. and Hrobjartsson, Asbjorn and Roberts, Chris and Shoukri, Mohamed and Streiner, David L.},
  title   = {Guidelines for Reporting Reliability and Agreement Studies ({GRRAS}) Were Proposed},
  journal = {Journal of Clinical Epidemiology},
  year    = {2011},
  volume  = {64},
  number  = {1},
  pages   = {96--106},
  doi     = {10.1016/j.jclinepi.2010.03.002},
  url     = {https://doi.org/10.1016/j.jclinepi.2010.03.002}
}

@article{dunivin2025scaling,
  author       = {Zackary Okun Dunivin},
  title        = {Scaling hermeneutics: a guide to qualitative coding with LLMs for
                  reflexive content analysis},
  journal      = {{EPJ} Data Sci.},
  volume       = {14},
  number       = {1},
  pages        = {28},
  year         = {2025},
  url          = {https://doi.org/10.1140/epjds/s13688-025-00548-8},
  doi          = {10.1140/EPJDS/S13688-025-00548-8},
  bibsource    = {dblp computer science bibliography, https://dblp.org}
}

@article{wen2026leveraging,
  author  = {Wen, Chuanchi and Clough, Paul and Paton, Rachel and Middleton, Rebecca},
  title   = {Leveraging Large Language Models for Thematic Analysis: A Case Study in the Charity Sector},
  journal = {AI \& Society},
  year    = {2026},
  volume  = {41},
  pages   = {731--748},
  doi     = {10.1007/s00146-025-02487-4},
  url     = {https://doi.org/10.1007/s00146-025-02487-4}
}

@inproceedings{dai2023llminloop,
  author    = {Dai, Shih{-}Chieh and Xiong, Aiping and Ku, Lun{-}Wei},
  title     = {{LLM}-in-the-loop: Leveraging Large Language Model for Thematic Analysis},
  booktitle = {Findings of the Association for Computational Linguistics: {EMNLP} 2023},
  year      = {2023},
  pages     = {9993--10001},
  address   = {Singapore},
  publisher = {Association for Computational Linguistics},
  doi       = {10.18653/v1/2023.findings-emnlp.669},
  url       = {https://doi.org/10.18653/v1/2023.findings-emnlp.669}
}

@article{terlecky2025llmcode,
  author  = {Han, Zhiyong and Tavasi, Aaron and Lee, JuYoung and Luzuriaga, Joshua and Suresh, Kevin and Oppenheim, Michael and Battaglia, Fortunato and Terlecky, Stanley R. and others},
  title   = {Can Large Language Models Be Used to Code Text for Thematic Analysis? An Explorative Study},
  journal = {Discover Artificial Intelligence},
  year    = {2025},
  volume  = {5},
  pages   = {171},
  doi     = {10.1007/s44163-025-00441-3},
  url     = {https://doi.org/10.1007/s44163-025-00441-3}
}

@misc{he2021debertav3,
  author        = {He, Pengcheng and Gao, Jianfeng and Chen, Weizhu},
  title         = {DeBERTaV3: Improving DeBERTa using ELECTRA-Style Pre-Training with Gradient-Disentangled Embedding Sharing},
  year          = {2021},
  month         = nov,
  eprint        = {2111.09543},
  archivePrefix = {arXiv},
  primaryClass  = {cs.CL},
  doi           = {10.48550/arXiv.2111.09543},
  url           = {https://arxiv.org/abs/2111.09543},
  note          = {arXiv:2111.09543v4 (revised 2023-03-24)},
  pages         = {1--16},
  numpages      = {16}
}

@article{corbin1990grounded,
  title   = {Grounded Theory Research: Procedures, Canons, and Evaluative Criteria},
  author  = {Corbin, Juliet M. and Strauss, Anselm},
  journal = {Qualitative Sociology},
  volume  = {13},
  number  = {1},
  pages   = {3--21},
  year    = {1990},
  doi     = {10.1007/BF00988593}
}

@book{saldana2011fundamentals,
  title     = {Fundamentals of Qualitative Research},
  author    = {Salda{\~n}a, Johnny},
  year      = {2011},
  publisher = {Oxford University Press},
  address   = {New York, NY},
  isbn      = {9780199737956}
}

@article{brown2020language,
  title={Language models are few-shot learners},
  author={Brown, Tom and Mann, Benjamin and Ryder, Nick and Subbiah, Melanie and Kaplan, Jared D and Dhariwal, Prafulla and Neelakantan, Arvind and Shyam, Pranav and Sastry, Girish and Askell, Amanda and others},
  journal={Advances in neural information processing systems},
  volume={33},
  pages={1877--1901},
  year={2020}
}

@misc{yang2025qwen3,
  title         = {Qwen3 Technical Report},
  author        = {An Yang and Anfeng Li and Baosong Yang and Beichen Zhang and Binyuan Hui and Bo Zheng and Bowen Yu and Chang Gao and Chengen Huang and Chenxu Lv and others},
  year          = {2025},
  eprint        = {2505.09388},
  archivePrefix = {arXiv},
  primaryClass  = {cs.CL},
  doi           = {10.48550/arXiv.2505.09388},
  url           = {https://arxiv.org/abs/2505.09388}
}

@article{aizawa2003information,
  title={An information-theoretic perspective of tf--idf measures},
  author={Aizawa, Akiko},
  journal={Information Processing \& Management},
  volume={39},
  number={1},
  pages={45--65},
  year={2003},
  publisher={Elsevier}
}

@inbook{sovrano2025magix,
  author="Sovrano, Francesco
and Miller, Tim",
title="MAGIX: A Unified Framework for the Use of XAI in Enterprises",
bookTitle="Enterprise AI",
year="2025",
publisher="Springer Nature Switzerland",
address="Cham",
pages="183--209",
isbn="978-3-032-01940-0",
doi="10.1007/978-3-032-01940-0_6",
url="https://doi.org/10.1007/978-3-032-01940-0_6"
}

%

\end{document}